\documentclass[preprint,12pt]{elsarticle} 
 
\usepackage{amssymb}
\usepackage{amsmath}

\usepackage{lineno}

\newenvironment{VarDescription}[1]%
  {\begin{list}{}{%
    \settowidth{\labelwidth}{\hspace{0cm}\textbf{#1}:}%
    \setlength{\leftmargin}{\labelwidth}\addtolength{\leftmargin}{\labelsep}}}%
  {\end{list}}

\usepackage{amsthm}
\theoremstyle{definition}
\newtheorem{definition}{Definition}[section]

\usepackage[algoruled,linesnumbered,vlined]{algorithm2e}

\usepackage{tikz}
\usepackage{pgfplots}
\usepackage{subfigure} 
\usepackage{pgfgantt}
\usepackage{multirow}
\usepackage{cases}
\renewcommand{\vec}[1]{\mathbf{#1}}

\pgfplotsset{compat=1.14}

\journal{Computers \& Operations Research}

\begin{document}

\begin{frontmatter}



\title{Strong Bounds for Resource Constrained Project Scheduling: Preprocessing and Cutting Planes}


\author[mymainaddress,myaddress1,myaddress4]{Janniele A. S. Araujo\corref{mycorrespondingauthor}}
\cortext[mycorrespondingauthor]{Corresponding author}
\ead{janniele@ufop.edu.br}
\author[mymainaddress,myaddress5]{Haroldo G. Santos}
\author[myaddress2,myaddress4]{Bernard Gendron}
\author[myaddress3,myaddress4]{Sanjay Dominik Jena}
\author[mymainaddress,myaddress1]{Samuel S. Brito}
\author[mymainaddress]{Danilo S. Souza}

\address[mymainaddress]{Department of Computing, DECOM, Universidade Federal de Ouro Preto}
\address[myaddress1]{Department of Computing and Systems, DECSI, Universidade Federal de Ouro Preto}
\address[myaddress4]{Interuniversity Research Centre on Enterprise Networks, Logistics and Transportation, CIRRELT}
\address[myaddress5]{Computer Science Department, CODeS, KU Leuven}
\address[myaddress2]{Department of Computer Science and Operations Research, Universit\'e de Montr\'eal}
\address[myaddress3]{Department of Management and Technology, \'Ecole des Sciences de la Gestion, Universit\'e de Qu\'ebec \`a Montr\'eal}

\begin{abstract}
Resource Constrained Project Scheduling Problems (RCPSPs) without preemption are well-known $\mathcal{N}\mathcal{P}$-hard combinatorial optimization problems. A feasible RCPSP solution consists of a time-ordered schedule of jobs with corresponding execution modes, respecting precedence and resources constraints. In this paper, we propose a cutting plane algorithm to separate five different cut families, as well as a new preprocessing routine to strengthen resource-related constraints. New lifted versions of the well-known precedence and cover inequalities are employed. At each iteration, a dense conflict graph is built considering feasibility and optimality conditions to separate cliques, odd-holes and strengthened Chv\'atal-Gomory  cuts. The proposed strategies considerably improve the linear relaxation bounds, allowing a state-of-the-art mixed-integer linear programming solver to find provably optimal solutions for 754 previously open instances of different variants of the RCPSPs, which was not possible using the original linear programming formulations.
\end{abstract}


\end{frontmatter}



\section{Introduction}
\label{sec:introduction}

This paper proposes automatic Mixed-Integer Linear Programming (MILP) reformulation strategies for non-preemptive Resource Constrained Project Scheduling Problems (RCPSPs) and its variants. From a theoretical point of view, RCPSPs are challenging combinatorial optimization problems, classified as $\mathcal{NP}$-hard \cite{Blazewicz1983,Garey1979}. These problems cover a wide range of applications, such as particle therapy for cancer treatment in healthcare \cite{Riedler2017},  civil engineering \cite{Liu2008}, manufacturing and assembly of large products \cite{Liu2014}, as well as development and launching of complex systems \cite{Deumeulemeester2002}. Comprehensive reviews on RCPSPs can be found, for example, in \citep{Artigues2008,Deumeulemeester2002,Schwindt2015}.

In this paper, we consider the following problem variants of the RCPSP, from the most specific one to the most generalized version:
\begin{description}
	\item[SMRCPSP:] single-mode resource-constrained project scheduling problem;
    \item[MMRCPSP:] multi-mode resource-constrained project scheduling problem;
    \item[MMRCMPSP:] multi-mode resource-constrained multi-project scheduling problem.
\end{description}

The SMRCPSP is the simplest variant and involves only one processing mode for each job. A feasible solution consists in the assignment of jobs at specific time periods over a planning horizon, respecting precedence and resource usage constraints. The resources in the SMRCPSP are renewable at each time period.

In the MMRCPSP, it is possible to choose between different job processing modes, each of them having different durations and consuming different amounts of resources. Two types of resources are available: the renewable resources at each time period and the non-renewable ones available for the entire project execution. While the use of renewable resources only impacts the delay/speedup of the projects, the use of non-renewable resources can produce infeasible solutions. 

A generalization of the previous problem variants to handle multiple projects and global renewable resources is the MMRCMPSP. While the previous two problem variants, in their objective (minimization) functions, consider the makespan, i.e., the total length of the schedule to finish all jobs, an important modeling feature of this generalization is an additional objective: the project delays from the Critical Path Duration (CPD). The CPD is a theoretical lower bound on the earliest finishing time ($\breve{e}_p^f$) of project $p$, computed by the Critical Path Method (CPM) \cite{Kelley1959} and disregarding any resource constraints.

\begin{figure}[!ht]
	\begin{center}
                \includegraphics[scale=0.77]{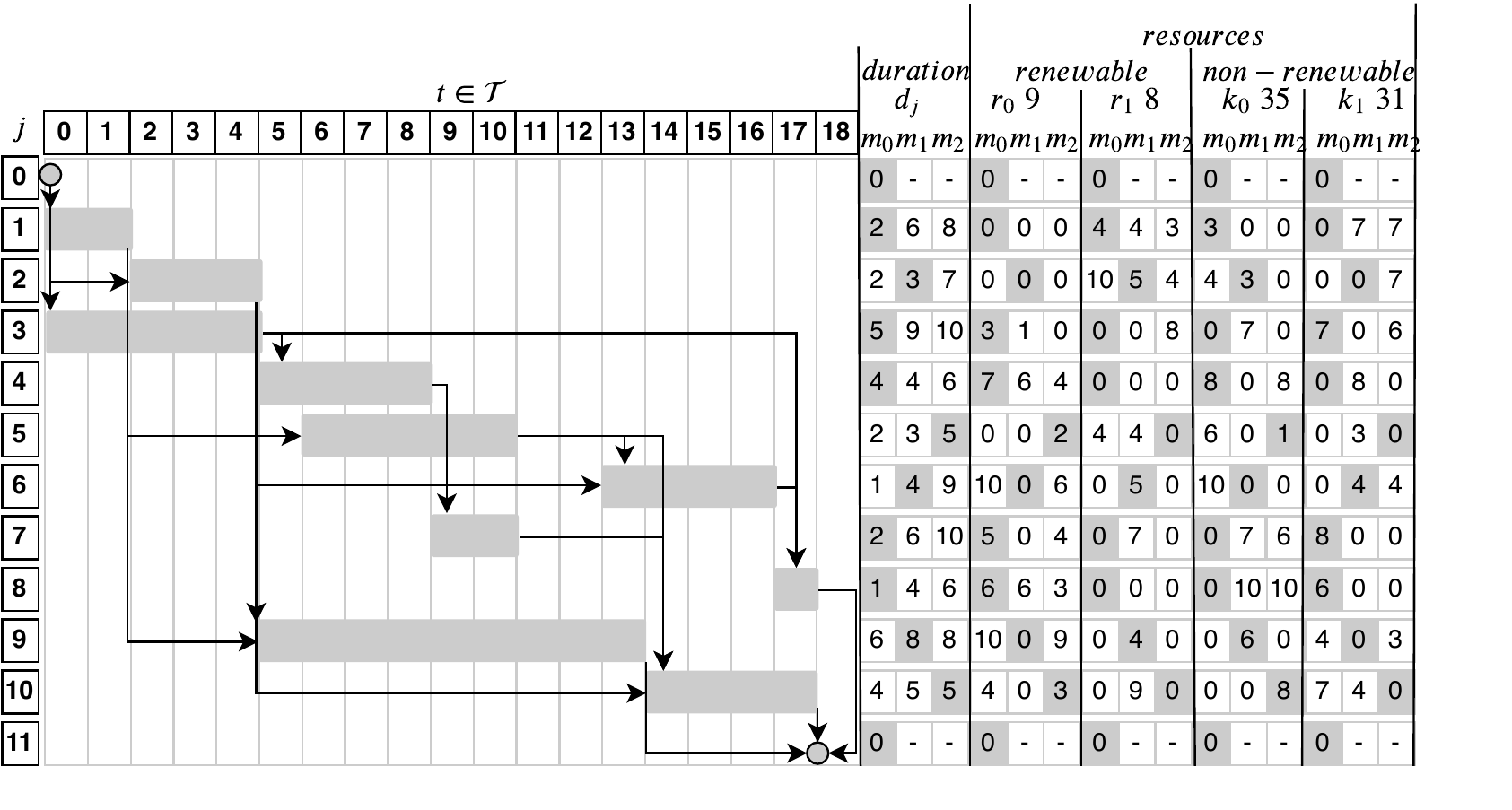}
               \caption{An optimal solution for instance $j102\_4.mm$ and its characteristics}
              \label{fig:gantt}
 \end{center}
\end{figure}

Figure \ref{fig:gantt} illustrates the example instance $j102\_4.mm$\footnote{http://www.om-db.wi.tum.de/psplib/files/j10.mm.tgz}, a small instance for the MMRCPSP variant. The figure shows the characteristics of this instance and one optimal solution. We will refer to this example instance to explain the cutting planes presented in the next sections. 

The instance illustrated in Figure \ref{fig:gantt} has twelve jobs. The first and the last are artificial jobs representing the start and the end of the project. Each job $j$ can be processed in three different modes $m$. Artificial jobs $0$ and $11$ have only one execution mode that does not consume resources and have duration $0$; thus columns $ m_1 $ and $ m_2 $ are filled with ``-". Two renewable resources $\{r_0,r_1\}$ with capacities $\{9,8\}$ and two non-renewable resources $\{k_0,k_1\}$  with capacities $\{35,31\}$ are available. This figure provides information about resource consumption and duration of job $j$ processing on mode $m$. The Gantt chart in the figure shows the starting time allocation and duration of jobs for the optimal solution found for this single project. Arcs represent the precedence relationship between jobs. Values emphasized in grayscale represent the active modes for this solution. The CPD corresponds to value $15$. Note that the earliest starting time ($\breve{e}_j^s$) of job $5$ corresponds to $2$, but due to the availability of resources at this time period, its allocation had to be postponed. The makespan is $18$ time units.

In this paper, we propose a new preprocessing technique to improve resource-related constraints by strengthening their coefficients using known information about different renewable, non-renewable resources and precedence constraints. We also propose a cutting plane algorithm employing five different cut separation algorithms. Conflict-based cuts such as cliques and odd-holes are generated considering an implicit dense conflict graph, which is updated at each iteration considering optimality and feasibility conditions. This conflict graph is also used in a MILP separation of the Chv\'atal-Gomory cuts to produce stronger inequalities. Furthermore, new lifted versions of the well-known disaggregated precedence cuts and cover cuts are separated. The contribution on the linear relaxation bound provided by each cut family is examined on experiments on a large set of benchmark instances of the three problem variants considered in this paper. Overall, stronger bounds were obtained, allowing a state-of-the-art MILP solver to find provably optimal solutions for 755 previously open instances, which was not possible using the original linear programming formulations. 

The paper is organized as follows. Section \ref{sec:literature}, reviews the literature related to our work. Section \ref{sec:exactmodel} introduces the integer programming formulation used along with some simple input data preprocessing routines based on upper bounds. The new strengthening procedure for renewable resources constraints is also presented in this section. Section \ref{sec:method} outlines the proposed cutting plane method, followed by a detailed description of each cut generation routine. In Section \ref{sec:results}, the computational results are presented. Finally, in Section \ref{sec:conclusion}, we conclude our work and discuss future research directions.

\section{Literature Review}
\label{sec:literature}
Various MILP based formulations have been proposed in the literature to model resource constrained project scheduling problems. \citet{Pritsker1969} proposed the first binary programming formulation where variables $x_{jt}$ indicate whether a job $j$ ends at time $t$ ($x_{jt} = 1$) or not ($x_{jt} = 0$). This formulation is known as the discrete-time or time-indexed formulation. The number of binary decision variables in this formulation is related to an upper bound  $\bar{t}$ for the number of time periods required to complete the project. Thus the number of variables is $\mathcal{O}(n\times\bar{t})$, where $n$ is the number of jobs. \citet{Kolisch1996} extended this formulation to handle different execution modes, adding one additional index $m$ to the binary variables and incorporating this index in the resource-related constraints. In \citet{Kone2011}, a new formulation was proposed based on events called on/off event-based (OOE) with $\mathcal{O}(n^2)$ variables.

Time-indexed formulations have been extensively studied and applied to the RCPSP (see, for example, \cite{Baptiste2004,CHRISTOFIDES1987262,Demassey2005,Hardin2008,Sankaran1999,deSouza1997}). Six time-indexed formulations were studied in \citet{Artigues2017}. These formulations are initially categorized into three groups according to the meaning of variables, as follows:

\begin{description}
	\item[Pulse:] pulse discrete time - PDT (the most used) formulation with binary variables $x_{jt} \ \forall j \in \mathcal{J}, \forall t \in \mathcal{T}$, such that $x_{jt} = 1$ if job $j$ starts at time $t$, otherwise $x_{jt} = 0$;
    \item[Step:] step discrete time - SDT formulation with binary variables $y_{jt} \ \forall j \in \mathcal{J}, \forall t \in \mathcal{T}$, such that  $y_{jt} = 1$ if job $j$ starts at time $t$ or before, otherwise  $y_{jt} = 0$;
    \item[On/Off:] on/off discrete time - OODT (the lesser used) formulation with binary variables $z_{jt} \ \forall j \in \mathcal{J}, \forall t \in \mathcal{T}$ such that $z_{jt} = 1$ if job $j$ is processed at time $t$, otherwise $z_{jt} = 0$.
\end{description}

For each one of these categories, it is possible to model the precedence constraints in a weak or strong way according to the linear programming (LP) relaxation strength.

\begin{description}
	\item[Aggregated:] this is a weak way to model precedence constraints, using variable coefficients greater than or equal to $1$; aggregated constraints generate $\mathcal{O}(n^2)$ inequalities, each one 
with $\mathcal{O}(\bar{t}\times m)$ variables, where $m$ is the maximum number of modes;
    \item[Disaggregated:] this is a stronger way to model precedence constraints, using variable coefficients equal to $\{-1,1\}$; disagregated constraints generate $\mathcal{O}(n^2\times\bar{t})$  inequalities, 
each one with $\mathcal{O}(\bar{t}\times m)$ variables.
\end{description}

\citet{Artigues2017} concluded that PDDT (pulse disaggregated discrete time), SDDT (step disaggregated discrete time) and OODDT (on/off disaggregated discrete time), which are formulations with disaggregated constraints, are all equivalent in terms of the strength of their LP-relaxations and belong to the family of strong time-indexed formulations. He also concluded that the formulations with their aggregated counterparts, PDT, SDT, and OODT, belong to a family of weak formulations and are also all equivalent in terms of their LP relaxation. 

Even though the aggregated constraints are weaker, they are much less dense. For this reason, papers from the literature (see, for example, \cite{CHRISTOFIDES1987262,Zhu2006}), begin the formulation with the aggregated constraints and add the disaggregated ones as cutting planes. As in previous works, our algorithm starts with the weak time-indexed formulation pulse discrete time (PDT) based on formulations proposed in \cite{Kolisch1996,Pritsker1969} for the MMRCMPSP version. In the next paragraphs, we review some computational approaches to handle MILP formulations for the RCPSPs.

Most of the exact algorithms for the RCPSPs \cite{Brucker1998272,Ripon2015,CHRISTOFIDES1987262,Demeulemeester1992} are built upon LP based Branch and Bound (B\&B) \cite{CHRISTOFIDES1987262,Land2010} algorithms. A key component in the design of these algorithms is which formulation is employed: compact formulations, with a polynomial number of variables and constraints, are usually able to quickly provide valid lower bounds since their LP relaxations are easily solved. The LP relaxation bounds are relatively weak but can be significantly improved by adding cutting planes in a Branch-and-Cut (B\&C) algorithm.

Cutting planes and strengthening constraints in MILP models are explored for RCPSP and to other variants of scheduling problems. \citet{Applegate1991} proposed cliques cuts, half cuts, and other cuts specific to job-shop scheduling problems (JSSP). \citet{Hardin2008} proposed a lifting procedure to cover-clique inequalities to a resource-constrained scheduling problem with uniform resources (URCSP) requirement. \citet{Cavalcante2001} applied cover cuts to the labor constrained scheduling problem (LCSP), based on practical requirements arising in industry.

\citet{Sankaran1999} proposed a cutting-plane algorithm for the SMRCPSP with minimal cover inequalities and clique inequalities to test problems provided by \citet{Patterson1974}. Also, they introduced three preprocessing techniques: reduction of lower and upper bounds; the identification of redundant constraints between resource and precedence constraints, and the coefficient strengthening in constraints \cite{Johnson1985}.

\citet{CHRISTOFIDES1987262} proposed a B\&B algorithm that uses disjunction arcs to handle resource conflicts. Four lower bounds are examined: the first one is based on the longest path in the precedence graph; the second one is based on an LP relaxation strengthened with cuts; the third one is based on  Lagrangian relaxation and the fourth one is based on disjunctive arcs. Their LP-based method incorporated the dynamic inclusion of disaggregated precedence constraints and resource-based conflict constraints for pairs of jobs. The first lower bound, based on precedence constraints, can be computed quickly and was used in the B\&B algorithm with other bounds. The second lower bound was promising, but was not used within the B\&B method. The Lagrangian relaxation technique was found to provide less competitive results. They provide additional inequalities for the time-indexed version. Finally, the fourth lower bound performed quite well, especially for problems with tight resource constraints. For the SMRCPSP, \citet{CHRISTOFIDES1987262} were able to prove the optimality of instances involving up to $25$ jobs and $3$ resources. 
 
\citet{Zhu2006} presented a B\&C algorithm for the MMRCPSP, including cuts derived from resource conflicts, where all resource constraints are in the form of generalized upper bound (GUB) constraints. Besides, disaggregated cuts from the precedence relationship for pairs of jobs $(j,s)$ in the precedence graph are included. To speedup the solution process, an adaptive branching scheme is developed along with a bound adjustment scheme that is always executed iteratively after branching. To optimize the solutions found in the first stage, the authors use a high-level neighborhood search strategy called Local Branching \cite{Fischetti2007}. For the MMRCPSP, the authors were able to prove the optimality $554$ of instances with 20-jobs and $506$ instances with 30-jobs.

Different techniques such as Constraint Programming (CP) and Satisfiability Solving (SAT) have also been used to solve the MMRCPSP. In this context, \citet{Demassey2005} uses CP techniques to provide valid inequalities to strengthen LP relaxations. A recent work using cutting planes as valid clauses for SAT is presented in \citet{Schnell2017}. They propose three formulations based on Constraint Programming to solve the MMRCPSP, using the G12 CP platform and the Solving Constraint Integer Programs (SCIP) as an optimization framework, both making use of solution techniques combining CP and SAT. They further combine MILP by inserting a new global constraint on the domain of renewable resources for SCIP. The authors achieved the same results with better computational times than \citet{Zhu2006}. They also were able to prove the optimality of $1428$ instances with 50-jobs and 100-jobs and to improve various lower and upper bounds.

\section{Integer Programming Formulation}
\label{sec:exactmodel}  

This section introduces the time-indexed formulation based upon the discrete time formulations proposed in \cite{Kolisch1996,Pritsker1969}. The most common objective function for the RCPSP is the makespan \cite{Artigues2008,Deumeulemeester2002,Kolisch1996,Pritsker1969,Zhu2006} minimizing the total schedule duration required to finish all jobs. \citet{Wauters2016} proposed a hierarchical objective to minimize the Total Project Delay (TPD) and the total makespan (TMS). The former, denoted formally in Eq.(\ref{eq:tpd}), is the main objective. The latter, given in Eq.(\ref{eq:tms}), is a tiebreaker.

\begin{equation}
\label{eq:tpd}
TPD = \sum_{p \in P}{PD_p}
\end{equation}

\begin{equation}
\label{eq:tms}
TMS = \max_{p\in \mathcal{P}}{f_p} - \min_{p\in \mathcal{P}}{\sigma_p}
\end{equation}

The project delay ($PD_p$) for a project $p \in \mathcal{P}$ is defined as the difference between its $CPD_p$ (which does not consider any resource constraints), and its makespan $MS_p =  f_p -\sigma_p$, the actual project duration taking into consideration resource constraints, the finishing time $f_p$ and the release date $\sigma_p$ of this project.  

\begin{equation}
PD_p = MS_p - CPD_p
\end{equation}

To work with all three problem variants in a unified way, we always report our results considering the TPD, which is generic enough to handle all problem variants considered in this paper.  

\subsection{Input Data}
\label{subsec:id}  

The following notation is used throughout this paper to describe the input data:

\begin{VarDescription}{xxxxxxx}
    \item [{$\mathcal{P}$}] set of all projects;
    \item [{$\mathcal{J}$}] set of all jobs;
	\item [{$\mathcal{M}_{j}$}] set of modes available for job $j \in \mathcal{J}$;
    \item [{$\mathcal{J}_{p}$}] set of jobs belonging to project $p$, such that $\mathcal{J}_{p} \subseteq \mathcal{J} \ \forall p \in \mathcal{P}$;
    \item [{$\mathcal{K}$}] set of non-renewable resources;
    \item [{$\mathcal{R}$}] set of renewable resources;
    \item [{$\mathcal{S}$}] set of direct precedence relationships between two jobs $(j, s) \in \mathcal{J} \times \mathcal{J}$;
    \item [{$\mathcal{T}\subset\mathbb{Z}^{+}$}] set of time periods in the planning horizon for all projects $p \in \mathcal{P}$;
    \item [{$\mathcal{T}_{jm}\subset\mathcal{T}$}] time horizon for each job $j \in \mathcal{J} $ on mode $m \in \mathcal{M}_{j}$, defined after preprocessing;
    \item [{$d_{jm}\in\mathbb{Z}^{+}$}] duration of job $j \in \mathcal{J}$ on mode $m \in \mathcal{M}_{j}$;
    \item [{$q_{kjm}\in\mathbb{Z}^{+}$}] required amount of non-renewable resource $k \in \mathcal{K}$ to execute job $j \in \mathcal{J}$ on mode $m \in \mathcal{M}_{j}$;
    \item [{$q_{rjm}\in\mathbb{Z}^{+}$}] required amount of renewable resource $r \in \mathcal{R}$ to execute job $j \in \mathcal{J}$ on mode $ m \in \mathcal{M}_{j}$;
    \item [{$\breve{q}_{k}\in\mathbb{Z}^{+}$}] available amount of non-renewable resource $k \in \mathcal{K}$; 
    \item [{$\breve{q}_{r}\in\mathbb{Z}^{+}$}] available amount of renewable resource $r \in \mathcal{R}$;
    \item [{$\sigma_p\in\mathcal{T}$}] release date of project $p$;
    \item [{$a_p\in\mathcal{J}_{p}$}] artificial job belonging to project $p \in \mathcal{P}$, which represents the end of the project.
\end{VarDescription}

\subsection{Preprocessing Input Data}
\label{subsec:pre}  

An effective way to reduce the search space is by identifying tight time windows in which it is valid to process jobs. A basic technique to define the earliest starting time $\breve{e}_j^s$ for jobs $j \in \mathcal{J}$ consists of computing the CPD using CPM \cite{Kelley1959} without considering resource constraints.  This methods allows to compute the $\breve{e}_j^s$ of all jobs, taking into consideration the precedence relationships. The longest path of a project, also known as the critical path, provides a lower bound for the completion time of each project.

Consider, for each project $p \in \mathcal{P}$, the release date $\sigma_p$, and lower bound based (i.e., the length of the critical path) $\lambda_p$ as input data and the value $\beta_p$, an upper bound for each project $p$, obtained from any feasible solution. Optimality conditions can be used to restrict the set of valid time periods when a job can be allocated. We initially consider the value  $\alpha$ computed by Eq.(\ref{eq:alpha}), that represents an upper bound to the maximum total project delay allowed.
\begin{equation}
\label{eq:alpha}
\alpha = \sum_{p\in \mathcal{P}}{\left(\beta_p-\sigma_p-\lambda_p\right)}
\end{equation}
 Thus, the maximum time period $\breve{t} \in \mathcal{T}$ that needs to be considered int the planning, can be obtained by Eq.(\ref{eq:MaxT}).
\begin{equation}
\label{eq:MaxT}
\begin{split}
 \breve{t} = \max_{p \in P}{ (\sigma_p+\lambda_p+\alpha) } \\
\mathcal{T} = \{ 0,...,\breve{t}\}
 \end{split}
\end{equation}

Analogously, upper bounds can be computed for processing times of jobs. The upper bounds can be strengthened if the selection of modes with different durations is also considered. The upper bounds are used, along with the duration of each job and without considering the resource constraints, to define the latest starting times ($\breve{l}_j^s$) for jobs $j \in \mathcal{J}$. A job $j$ from a project $p$ when processed at mode $m$ will push forward (i.e., postpone) all successor jobs by exactly $d_{jm}$ time units. Consider set $\overline{\mathcal{S}}_j$, containing the entire chain of successors of job $j$ on the longest path from job $j$ to the artificial job $a_p$ (indicating the project completion). Let lower bound $\mathcal{L}_{jm}$ be the total duration in this path, computed considering only the fastest processing modes for each job in this chain. The maximum allocation time or latest starting time $(\breve{l}^{s}_{jm})$ for a job $j$ from a project $p$ when processing on mode $m$ to $\mathcal{T}_{jm}$ is given by Eq.(\ref{eq:MaxTJM}).
\begin{equation}
\label{eq:MaxTJM}
\begin{split}
  \breve{l}^{s}_{jm} = \sigma_p +\lambda_p -  \mathcal{L}_{jm} +\alpha \\
  \mathcal{T}_{jm} = \{  \breve{e}_{j}^{s},...,\breve{l}^{s}_{jm}\}
 \end{split}
\end{equation}
Similar bounds can be derived for any two jobs in this path also considering the fastest processing modes for all jobs except the first one: 

\begin{VarDescription}{xxxxxxx}
  \item [{$\breve{d}_{jms}$}] the shortest path in the precedence graph considering the length of the arcs between job $j$ and successor job $s \in \overline{\mathcal{S}}_j$ considering mode $m \in \mathcal{M}_j$;
    \item [{$\breve{d}^*_{js}$}] the shortest path in the precedence graph considering the length of the arcs between job $j$ and successor job $s \in \overline{\mathcal{S}}_j$  considering $j$ fastest mode.
    \label{id:d}
\end{VarDescription}

\subsection{Formulation}
\label{subsec:form}  

Binary decision variables are used to select the mode and starting times for the jobs. They are defined as follows: 

\begin{equation*}
    x_{jmt} =
        \begin{cases}
            \; 1 &
            \textrm{if job } j \in \mathcal{J} \textrm{ is allocated on mode } m \in \mathcal{M}_{j} \\& \textrm{at starting time } t \in \mathcal{T}_{jm};\\
            \; 0 & \textrm{otherwise}.
        \end{cases}
        \label{eq:eff}  
\end{equation*}

We introduce in this formulation on/off discrete time variables studied in \citet{Artigues2017} to allow resources constraints and cutting planes, detailed in the next sections, to be expressed with fewer variables. The following binary decision variables indicate during which time periods jobs are being processed:
\begin{equation*}
    \label{eq:eff2}  
    z_{jmt} =
        \begin{cases}
            \; 1 &
            \textrm{if the  job } j \in \mathcal{J} \textrm{ is allocated on mode } m \in \mathcal{M}_{j} \textrm{ and} \\ & \textrm{   is being processed during time } t \in \mathcal{T}_{jm};\\
            \; 0 & \textrm{otherwise}.
        \end{cases}
\end{equation*}

The objective function minimizes the total project delay over the project completion times for projects and their critical paths. Consider the following integer variable included in the objective function:
\begin{VarDescription}{xxxxxxx}
\item [{$h \in \mathbb{Z}^{+}$}] integer variable used to compute the makespan, included in the objective function with a small coefficient $\epsilon$ to break ties.
\end{VarDescription}

\noindent \textit{Minimize:}
\begin{eqnarray}
    \label{model:foCons}
\displaystyle  \sum_{p\in \mathcal{P}}\sum_{m\in \mathcal{M}_{a_p}}\sum_{t\in \mathcal{T}_{a_pm}}{ \left[t - (\sigma_p + \lambda_p)\right] x_{a_pmt} }  + \epsilon  h
\end{eqnarray}

\noindent \textit{subject to:}
\begin{eqnarray}
    \label{model:maximoUmaVez} 
   \sum_{m\in \mathcal{M}_{j}}\sum_{t \in \mathcal{T}_{jm}}{ x_{jmt} }
         = 1 \ \ \forall j\in \mathcal{J}   \\ [0.5cm]
    \label{model:recursoNR}
        \sum_{j\in \mathcal{J}} \sum_{m\in \mathcal{M}_{j}} \sum_{t \in \mathcal{T}_{jm}} q_{kjm} x_{jmt}
         \leq  \breve{q}_{k}
\ \ \forall k\in \mathcal{K} \\[0.5cm]
    \label{model:recursoR}
          \sum_{j\in \mathcal{J}}\sum_{m\in \mathcal{M}_{j}}
         { q_{rjm}  z_{jmt}}
        \leq \breve{q}_{r}
\ \ \forall r \in \mathcal{R}, \forall t \in \mathcal{T} 
        \end{eqnarray}
\begin{eqnarray}
\label{model:precedencia1}
 \sum_{m\in \mathcal{M}_{j}}\sum_{t \in \mathcal{T}_{jm}}{\left(t+d_{jm}\right)  x_{jmt}}\ -  \ \sum_{z\in \mathcal{M}_{s}}\sum_{i \in \mathcal{T}_{sz}}{i  x_{szi}}
	  \leq  0 \nonumber \\
 \forall j\in \mathcal{J}, \forall s\in \mathcal{S}_j \\ [0.5cm]
   \label{model:varlnkxz}
 z_{jmt} - \sum_{t^{'} = (t-d_{jm}+1)}^{t}{x_{jmt^{'}}}
         = 0  \ \  \forall j \in \mathcal{J}, \forall m \in \mathcal{M}_j, \forall t \in \mathcal{T}_{jm} \\ [0.5cm]
   \label{model:ymax} 
   h - \sum_{m\in \mathcal{M}_{a_{p}}}\sum_{t \in \mathcal{T}_{a_{p}m}}{ t  x_{a_{p}mt} }
         \geq 0 \ \  \forall p \in \mathcal{P}\\ [0.5cm]
\label{model:varDecx}
x_{jmt}  \in  \{0,1\} \ \ \forall j \in \mathcal{J}, \forall m \in \mathcal{M}_j, \forall t \\ [0.5cm]
\label{model:varDecz}
z_{jmt}  \in  \{0,1\}  \ \forall j \in \mathcal{J}, \forall m \in \mathcal{M}_j, \forall t \in \mathcal{T}_{jm}   \\ [0.5cm]
\label{model:varInth}
h \geq 0  
\end{eqnarray}

Constraints (\ref{model:maximoUmaVez}) ensure that each job is allocated to exactly one starting time and one mode. Constraints (\ref{model:recursoNR}) and (\ref{model:recursoR}) are capacity constraints for non-renewable and renewable resources, respectively   . Constraints (\ref{model:precedencia1}) force precedence relationships to be satisfied.  Constraints (\ref{model:varlnkxz}) link variables $z$ and variables $x$.  Constraints (\ref{model:ymax}) compute the total makespan. Finally, constraints (\ref{model:varDecx}), (\ref{model:varDecz}) and (\ref{model:varInth}) respectively ensure that variables $x$ and $z$ can only assume binary values and $h$ can only assume nonnegative values.

\subsection{Preprocessing MILP Formulation}
\label{subsec:premip}  

\citet{Johnson1985} introduce an interesting preprocessing method to strengthen constraint coefficients using the knapsack structure of resource constraints. This preprocessing was used in \citet{Sankaran1999} for the SMRCPSP and analyzes one resource usage constraint at time. In this paper, we propose a preprocessing technique that considers various constraints (precedence and the usage of other renewable and non-renewable resources), besides the renewable resource constraint, which will be strengthened.

The proposed procedure to strengthen resource usage constraints (\ref{model:recursoR}) was inspired by Fenchel cutting planes \cite{Boyd1992,Boyd1994}. Fenchel cutting planes are based on the enumeration of incidence vectors to find the most violated inequality for a subset of binary variables. In our paper, we enumerate feasible subsets of jobs and modes to create a linear problem to find the best possible strengthening of a given resource constraint. 

The strengthening procedure is presented in Algorithm \ref{alg:strengthening}. First, it computes, for each $t$, a set $\mathcal{G}_t$ composed of all jobs and modes ($j,m$) available for processing at time $t \in \mathcal{T}_{jm}$ (see Algorithm \ref{alg:strengthening}, lines \ref{alg:t}--\ref{alg:gt}). Formally, these sets can be computed as stated in Eq.(\ref{eq:gt}).
\begin{equation}
\label{eq:gt}
     \mathcal{G}_t  = \{ \ (j,m) \ \in \ \mathcal{J}  \times \mathcal{M}_j \mid \  t \in \mathcal{T}_{jm} \}
\end{equation}

To illustrate the coefficient strengthening technique, consider for $t=4$, the jobs and modes that make up the set  $\mathcal{G}_4 = \{(1,0), (1,1), (1,2), (2,1), (2,2),\\ (3,0), (5,0), (5,1), (5,2), (6,1), (6,2), (9,1), (9,2)\}$. Consider also, the following original constraint restricting the usage of renewable resource $r_0$ at time $t=4$:
\begin{eqnarray*}
0 z_{1,0,4} + 0 z_{1,1,4} + 0 z_{1,2,4} + 0 z_{2,1,4} + 0 z_{2,2,4} + 3 z_{3,0,4} + 0 z_{5,0,4} + 0 z_{5,1,4} + \\ 2 z_{5,2,4} + 0 z_{6,1,4} + 6 z_{6,2,4} +0 z_{9,1,4} +  9 z_{9,2,4}  \leq 9.
\end{eqnarray*}

The next step is to enumerate all valid combinations of jobs and modes $(j,m)$ that can be processed in parallel at time $t$, i.e., that satisfy all resources constraints and do not have precedence relations among each other. \citet{Mingozzi1998} designated these valid combinations as \textit{feasible subsets}. 

Let $\mathcal{E}_t = (\bar{e}_1, \bar{e}_2, \ldots, \bar{e}_n)$ be the set of all these feasible subsets. This set can be computed by using a simple backtracking algorithm that recursively proceeds, from level $0$ to $|\mathcal{G}_{t}|$; tentatively fixing the allocation of each respective job and mode to 0 or 1; proceeding to the next level on the search space only when the partial fixation to the current level does not violate any resource or precedence constraint. 

If we enumerate all the possibilities over $\mathcal{G}_4$, we could have $2^{13}=8192$ feasible subsets. However, due to multiple constraints, there are only $51$ feasible subsets in $\mathcal{E}_4$:

   \begin{flushright}\footnotesize
\texttt{$\mathcal{E}_4 =$ [\{(1,0)\}, \fbox{\{(2,2),(3,0),(1,0)\}}, \{(2,2),(1,0)\}, \{(3,0),(1,0)\}, \{(1,1)\}, \{(2,2),(3,0),(1,1)\}, \{(2,2),(1,1)\}, \{(3,0),(1,1)\}, \{(1,2)\}, \{(2,1),(3,0),(1,2)\}, \{(2,1),(1,2)\}, \{(2,2),(3,0),(1,2)\}, \{(2,2),(1,2)\}, \{(3,0),(1,2)\}, \{(2,1)\}, \{(3,0),(5,2),(2,1)\}, \{(3,0),(2,1)\}, \{(5,2),(2,1)\}, \{(2,2)\}, \{(3,0),(5,0),(2,2)\}, \{(3,0),(5,1),(2,2)\}, \{(3,0),(5,2),(2,2)\}, \{(3,0),(2,2)\}, \{(5,0),(2,2)\}, \{(5,1),(2,2)\}, \{(5,2),(2,2)\}, \{(3,0)\}, \{(5,0),(9,1),(3,0)\}, \{(5,0),(3,0)\}, \{(5,1),(9,1),(3,0)\}, \{(5,1),(3,0)\}, \{(5,2),(9,1),(3,0)\}, \{(5,2),(3,0)\}, \{(6,1),(3,0)\}, \{(6,2),(9,1),(3,0)\}, \{(6,2),(3,0)\}, \{(9,1),(3,0)\}, \{(5,0)\}, \{(9,1),(5,0)\}, \{(9,2),(5,0)\}, \{(5,1)\}, \{(9,1),(5,1)\}, \{(9,2),(5,1)\}, \{(5,2)\}, \{(9,1),(5,2)\}, \{(6,1)\}, \{(9,2),(6,1)\}, \{(6,2)\}, \{(9,1),(6,2)\}, \{(9,1)\}, \{(9,2)\}]}.
    \end{flushright}
As an example of the impact of considering multiple constraints for reducing the number of valid incidence vectors, if we do not consider precedence constraints and we just consider one renewable resource constraint in the enumeration process, $182$ feasible subsets would be built.For the maximal feasible subset \texttt{\{(2,2),(3,0),(1,0)\}}, highlighted above inside the box, if we do not consider all resources and precedence constraints, it will be extended to \texttt{\{(2,2),(3,0),(5,2),(1,0)\},  \{(2,2),(3,0),(6,2),(1,0)\}}.

The $i^{th}$ feasible subset $\bar{e}_i$ contains ordered pairs $(j,m) \in \mathcal{G}_t$. For each renewable resource $r$ with capacity $c$ and time $t$ the following linear program ($W_{rt}$) can be used to strengthen constraints (\ref{model:recursoR}) if the enumeration process is successful, i.e., a pre-defined maximum number of iterations ($it$) was not reached, (see Algorithm     \ref{alg:strengthening}, lines \ref{alg:suc}--\ref{alg:replace}). Consider the continuous variables $u_{jm}$, indicating the number of consumed units of resource $r$ by job $j$ at mode $m$ in the strengthened constraint of the following linear programming  ($W_{rt}$ model):\\

 \noindent \textrm{\textit{ Maximize:}} 
 	\begin{eqnarray}
 \label{eq:miprein1}
     \displaystyle \sum_{(j,m) \in \mathcal{G}_t}  u_{jm}
    \end{eqnarray}
 \noindent \textrm{\textit{Subject  to:}} 
   \begin{eqnarray}
   \label{eq:miprein2}
    \displaystyle \sum_{(j,m) \in \bar{e}} u_{jm}  \leq c  \  \forall  \bar{e} \in \mathcal{E}_{t}   \\ [0.5cm]
   \label{eq:miprein3}
    q_{rjm} \leq u_{jm}  \leq c \ \forall (j,m) \in \mathcal{G}_t 
     \end{eqnarray}

Consider $\forall (r,t) \in \mathcal{R} \times \mathcal{T}: \bar{q}_{rjmt} = u^*_{jm}$ from $W_{rt}$, where $u^*_{jm} $ is the optimal solution in $W_{rt}$. This value is introduced as a new input data for the main formulation:

\begin{VarDescription}{xxxxxxx}
   \item [{$\bar{q}_{rjmt}$}] new values for required amount of renewable resource $r \in \mathcal{R}$ to execute job $j \in \mathcal{J}$ on mode $ m \in \mathcal{M}_{j}$ at time $t$.
\end{VarDescription}

Constraints (\ref{model:recursoRZ}) are created with the new values $\bar{q}_{rjmt}$, yielding improved capacity constraints for renewable resources.

\begin{eqnarray}
            \label{model:recursoRZ}
         \sum_{j\in \mathcal{J}}\sum_{m\in \mathcal{M}_{j}} \bar{q}_{rjmt} z_{jmt}
        \leq \breve{q}_{r}
      \ \    \forall r \in \mathcal{R}, \ \forall t \in \mathcal{T}
\end{eqnarray}

Due to the bounds on variables $u$,  constraints (\ref{model:recursoRZ}) always dominate the original constraints (\ref{model:recursoR}), since $\bar{q}_{rjmt} \geq q_{rjm}$. In particular, whenever $u_{jm} > q_{rjm}$ it dominates strictly. The following defines dominance between two generated cuts \cite{Wolsey1998}: 
\begin{definition}
 Let $ c^{1^{T}} x \leq r^{1}$ and $ c^{2^{T}} x \leq r^{2}$ be two inequalities. We say $ c^{1^{T}} x \leq r^{1}$ dominates  $ c^{2^{T}} x \leq r^{2}$ if $c_{i}^{1} \geq c_{i}^{2} \ \forall i$ and $r^{1} \leq r^{2}$; if at least one of these inequalities is satisfied as an strict inequality, then there is a strict dominance.
 \label{def:domination}
\end{definition}

An interesting property of this procedure is that it may strengthen constraints of resources that are \emph{not scarce}, given the scarceness of other resources and/or precedence constraints.

\paragraph{Example} By solving the MILP model $\mathcal{W}_{0,4}$ to our example introduced above, the coefficient of the emphasized variable $\boldsymbol{z_{5,2,4}}$ corresponding to job $5$ on mode $2$ processed at $t=4$ can be strengthened to $6$, without excluding any feasible integer solution. We can generate the following strengthened constraint:
\begin{eqnarray*}
0 z_{1,0,4} + 0 z_{1,1,4} + 0 z_{1,2,4} + 0 z_{2,1,4} + 0 z_{2,2,4} + 3 z_{3,0,4} + 0 z_{5,0,4} + 0 z_{5,1,4} + \\ \boldsymbol{6 z_{5,2,4}} +  0 z_{6,1,4} + 6 z_{6,2,4} + 0 z_{9,1,4} + 9 z_{9,2,4}  \leq 9.
\end{eqnarray*}

For the SMRCPSP and MMRCPSP, the original constraints (\ref{model:recursoR}) can be replaced by new constraints presented in (\ref{model:recursoRZ}) in the case that the latter are stronger. For the MMRCMPSP, the new constraints (\ref{model:recursoRZ}) are created to strengthen renewable resources for  each project separately, and the original constraints remain in the model.

We consider a time limit $tl$, which will be checked after the subset enumeration procedure for each $t$. If the time limit is reached, the remaining time periods $t$ will be skipped and the algorithm is terminated (see Algorithm \ref{alg:strengthening}, lines \ref{alg:time}--\ref{alg:timeend}).  We continue to find feasible subsets while it does not reach the last element of $\mathcal{G}_{t}$. If the maximum number of iterations $it$ is reached, we stop the process (by returning $\emptyset$, see Algorithm \ref{alg:strengthening}, lines \ref{alg:btsb}--\ref{alg:suc}) and continue to the next value of $t$ on the strengthening algorithm. 
\begin{algorithm}[!ht]\footnotesize
    \KwData{ RCPSP model $\mathcal{M}$, $it$ iteration limit, $tl$ time limit, set $\mathcal{J}$, set $\mathcal{M}_{j}$, set $\mathcal{T}$, set $\mathcal{R}$, set $\mathcal{K}$, set $\overline{\mathcal{S}}_j$}
   \For{ (each $t \in \mathcal{T}$)} { \label{alg:t}
        $\mathcal{G}_{t} \gets$ \texttt{compute\_by\_Eq.\ref{eq:gt}($\mathcal{J}$,$\mathcal{M}$,$t$)}; \label{alg:gt}\\ 
        $\mathcal{E}_{t} \gets \emptyset$; \ $ite \gets 0$;\\
        $\mathcal{E}_{t} \gets$ \texttt{subset\_bt($it$,$ite$,$tl$,$\mathcal{G}_{t}$,$\mathcal{R}$,$\mathcal{K}$,$\overline{\mathcal{S}}_j$,$\mathcal{E}_{t}$)}; \label{alg:btsb}\\
         \If{ ($\mathcal{E}_{t} \neq \emptyset$) }{\label{alg:suc}
            \For{ (each $r \in \mathcal{R}$ with capacity $c$)} {
                 $\mathcal{W}_{rt} \gets$ \texttt{create\_strengthening\_MIP($r$,$c$,$\mathcal{G}_{t}$,$\mathcal{E}_{t}$)};\\
                 $\bar{q}_{rjmt} \gets$; \texttt{opt($\mathcal{W}_{rt}$)};\\
             \texttt{create\_replace\_constraints(Ineq.(\ref{model:recursoR}), $\bar{q}_{rjmt}$, Ineq.(\ref{model:recursoRZ}))};\label{alg:replace}
            }\label{alg:endsuc}
         }
        \If{time limit $tl$ is reached}{\label{alg:time} \textbf{break};
        }\label{alg:timeend}
     }
    \caption{\texttt{strengthening\_procedure}}
    \label{alg:strengthening}
\end{algorithm} 
  
\section{The Cutting Plane Algorithm}
\label{sec:method}  

The performance of general purpose MILP solvers on a given formulation strongly depends on how tight is the LP relaxation (dual bound) to the optimal solution \cite{Wolsey1998}. Cutting planes are commonly used to improve this bound by iteratively adding violated cuts.

The proposed cutting plane algorithm uses traditional RCPSP cuts enhanced with new lifting techniques (see Subsections \ref{subsec:cover} and \ref{subsec:prece}, respectively for cover and precedence cuts), conflict-based cuts (see Subsection \ref{subsec:conflict} for clique and odd-holes cuts) and strengthened Chv\'atal-Gomory cuts (see Subsection \ref{subsec:cg}) generated from an implicit dense conflict dynamic graph. 

\begin{figure}[!ht]
        \begin{center}
 \includegraphics[width=0.9\linewidth]{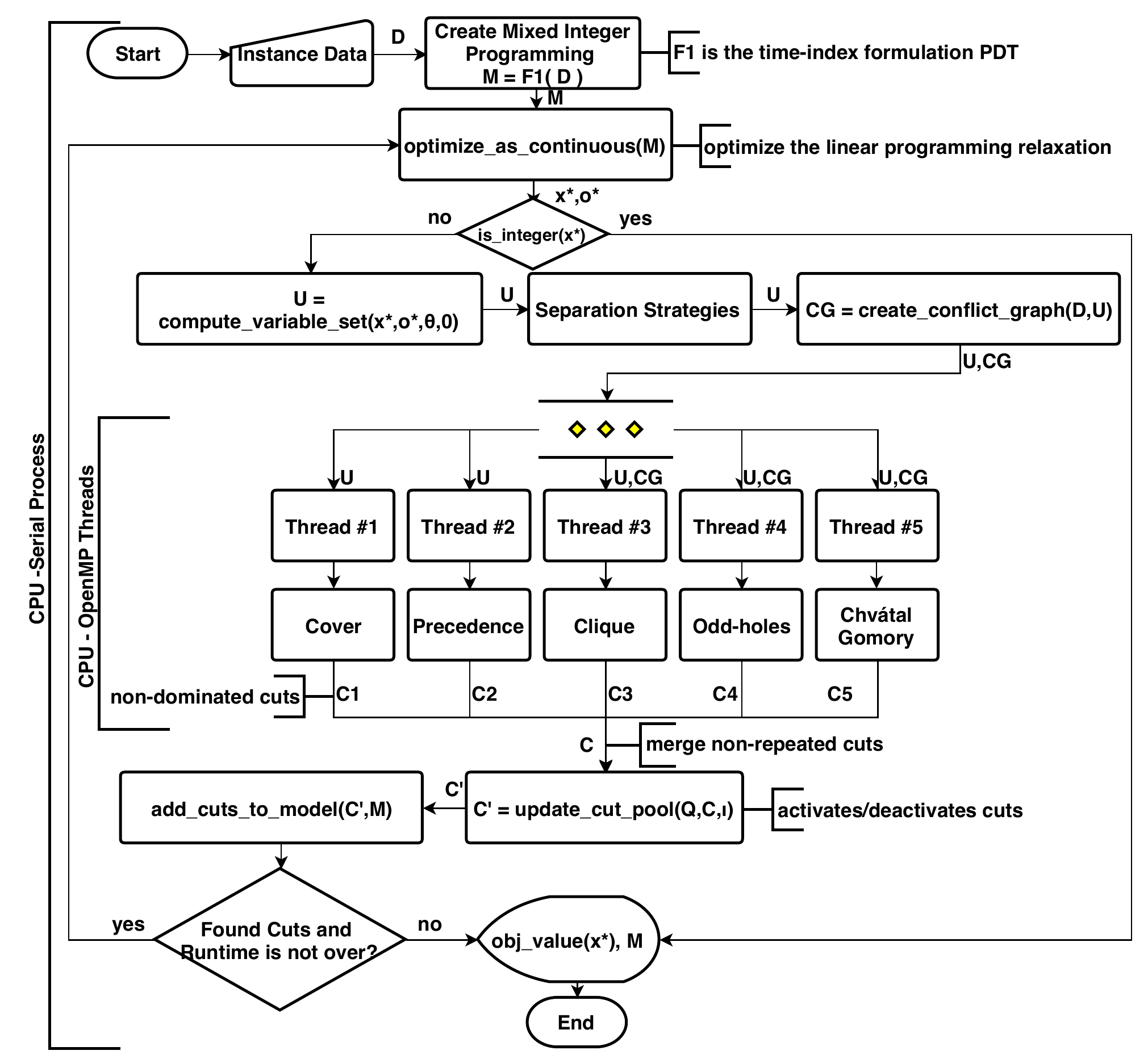}
                \caption{Execution flow of the proposed cutting plane method}
                \label{fig:outline}
       \end{center}
\end{figure}

An outline of the proposed method is depicted in Figure \ref{fig:outline}. After the instance data is read, we create the mathematical programming model (\textbf{M}) explained in Section \ref{sec:exactmodel} and execute the preprocessing routines. Then, the optimal LP relaxation is computed and if a fractional solution is obtained, different search methods are started to separate violated inequalities. Since the separation of these inequalities may involve the solution of $\mathcal{N}\mathcal{P}$-hard problems, we execute the separation procedures in parallel. This allows us to save some processing time in order to process a larger number of iterations in the cutting plane algorithm within the given time limit.

All generated cuts are inserted into a cut pool where repeated inequalities are discarded. Our algorithm quickly discards repeated cuts by using a hash table. While checking for repeated cuts is very fast, the dominance check is slower since it requires checking the contents of the cuts. We only check the dominance in the pool of cover separation since they generated less cuts compared with other types of cuts. 

If new cuts have been found after the separation procedure, a stronger formulation is obtained, and the process is repeated. When the time limit is reached or when no further cut is generated, the strengthened model and its objective function value are returned. In the following, we present the different inequalities that are separated as well as the algorithmic aspects involved in their separation.

\subsection{Lifted RCPSP Knapsack Cover Cuts - LCV}
\label{subsec:cover}

Resource usage constraints present a knapsack structure which was exploited in \cite{Zhu2006} to generate GUB cover (CV) cuts. Consider the general case of a constraint in the form $ \displaystyle \sum_{j \in \mathcal{N}} b_j   z_j \leq c,(b,c)\in {\mathbb{N}^{+}}^{n}\times \mathbb{N}^{+} $ for a set $\mathcal{N}$ of binary variables. The following knapsack problem can be solved to generate a valid inequality that cuts fractional point $z^*$:\\

\noindent \textit{Minimize}
	\begin{eqnarray}
	\label{eq.coverinit}
		\gamma (z^*)=\sum_{j \in \mathcal{N}} (1-z_{j}^{*}) v_j
    \end{eqnarray}
\noindent \textit{subject to:}	\begin{eqnarray}
    \sum_{j \in \mathcal{N}}{ b_j v_j} > c\\
    v_j\in \{0,1\} \forall j \in \mathcal{N}
    \label{eq.coverend}
    \end{eqnarray}

Whenever a solution with $\gamma (z^*) < 1$ is found involving a set $\mathcal{V}=\{j \in \mathcal{N} : v_j=1\}$ a violated inequality in the form $\displaystyle \sum_{j \in \mathcal{V}}z_j \leq |\mathcal{V}|-1$ is generated. Since only active variables $(z_j^*>0)$ are considered in this separation, many dominated inequalities can be generated in different iterations. 

While general purpose lifting strategies \cite{Gu2000} can be used, specific problem information can provide an effective procedure to produce lifted cover inequalities. A variable in a traditional cover inequality represents whether a job $j$ and mode $m$ were allocated or not at time $t$. The proposed lifted RCPSP knapsack cover cut (LCV) separation routine may include, for each job $j$, variables of additional processing modes without increasing the right-hand-side, producing much stronger cuts. In the single mode case the strengthening will be similar to the one obtained with traditional lifting (see \cite{Balas1975,Balas1978,Nemhauser1994}) techniques.

The new lifted knapsack cover separation problem for the RCPSP is solved for each renewable resource and each time period. Consider period $t^{*}$, resource $r$ with capacity $\breve{q}_{r}$ and a fractional solution with variable values $z_{jmt^*}^{*}$. The decision variables are:

\begin{VarDescription}{xxxxxxxxx}
    \item [{$v_{jm}\in\{0,1\}$}] if variable of job $j$ at mode $m$ is selected (1) or not (0);
    \item [{$o_{j}\in\{0,1\}$}] if at least one mode for job $j$ is included in the cut (1) or not (0);
    \item [{$\underline{w}_{jm}\in\{0,1\}$}] if job $j$ has $m$ as the selected mode with the smallest resource consumption from the selected ones (1) or not (0);
    \item [{$\underline{e}\in\mathbb{Z}^{+}$}] resource consumption excess if modes with smallest resource consumption are selected.
    \item [{$\overline{v}\in\mathbb{R}^{+}$}] cut violation.
\end{VarDescription}

The LCV separation problem is given by:\\

\noindent \textit{Maximize}
\begin{eqnarray}
    \label{model:foCons2}
\displaystyle  \omega  \overline{v} +\sum_{j\in \mathcal{J}}\sum_{m\in \mathcal{M}_{j}}\mu v_{jm}
\end{eqnarray}
\noindent \textit{subject to:}	\begin{eqnarray}
 &  & o_{j}=\sum_{m\in \mathcal{M}_{j}}\underline{w}_{jm}\ \ \ \forall j\in J\label{eq:lnkew}\\
 &  & \underline{e}=\sum_{j\in \mathcal{J}}\sum_{m\in \mathcal{M}_{j}}q_{rjm}  \underline{w}_{jm}-\breve{q}_{r}\label{eq:e2}\\
 &  & v_{jm} \leq \sum_{m'\in \mathcal{M}_{j}:q_{rjm'}\leq q_{rjm}}\underline{w}_{jm'}\ \ \ \forall j\in \mathcal{J},m\in \mathcal{M}_{j}\label{eq:w2}\\
 &  & \overline{v}=\sum_{j\in \mathcal{J}}\sum_{m\in \mathcal{M}}z_{jmt^*}^{*} v_{jm}-\sum_{j\in \mathcal{J}}o_{j}+1\label{eq:viol}\\
 &  & 0.005 \leq \overline{v} \leq \infty \label{eq:minViol}\\
 &  & 1 \leq \underline{e} \leq \infty \label{eq:qfeas2}\\
 &  & o_{j}, v_{jm}, \underline{w}_{jm} \in  \{0,1\} \ \ \ \forall j \in \mathcal{J}, \ \forall m \in \mathcal{M}_j \label{eq:bin}
\end{eqnarray}

The objective function (\ref{model:foCons2}) maximizes a hierarchical objective function composed of, first, the cut violation and, second, the inclusion of additional jobs and modes in the generated inequality to produce stronger cuts ($\omega \gg \mu$). These weights cannot be too large or too small to prevent numerical instability in the solvers. Constraints (\ref{eq:lnkew}) ensure that $o_{j}$ is only activated when some mode is selected for job $j$ as the mode with lower resource usage. Constraints (\ref{eq:e2}) and (\ref{eq:qfeas2}) ensure that a cover is produced. Constraints (\ref{eq:w2}) ensure that only modes with resource usage greater than or equal to the mode with the smallest resource usage selected are allowed for lifting. Equality (\ref{eq:viol}) computes the cut violation and constraint (\ref{eq:minViol}) ensures that a violated cut is produced. Finally,  constraints (\ref{eq:bin}) ensure that variables $o_j,v_{jm}$ and $\underline{w}_{jm}$ can only assume binary values.


Let $C$ be the set of selected jobs and modes (with $v_{jm}=1$) in the solution of the problem above. Further, let $o_j$ indicate its corresponding solution values. We may generate they following LCV cut:

\begin{equation}
\label{eq:selectedmodes}
\sum_{(j,m)\in C}z_{jmt}\leq\sum_{j\in J}o_{j}-1  \ \ \ \forall t \in \mathcal{T}
\end{equation}

For a valid cover cut the value on the right-hand-side would be the size of the set minus $1$, with the lifting strategy whenever it chooses more than one mode per job the sum of the right-hand-side will be strictly smaller than the size of the set minus $1$, generating a stronger cut as in the example below.

\paragraph{Example}  Consider the following cut for resource $r_0$ on $t=8$, and jobs \{4,8\} processing respectively on modes \{0,2\} from Figure \ref{fig:gantt}, generated using the separation described in (\ref{eq.coverinit})-(\ref{eq.coverend}):
\begin{eqnarray*}
(CV) = z_{4,0,8} + z_{8,2,8} \leq 1.
\end{eqnarray*} 

This cut can be lifted since job $8$ has other modes \{$0,1$\}, each of them consuming $6$ units of resource $r_0$, more than the current mode \{$2$\} that consumes $3$ units, forming still a valid inequality:
\begin{eqnarray*}
(LCV)= z_{4,0,8} + \boldsymbol{z_{8,0,8} + z_{8,1,8}} + z_{8,2,8} \leq 1.
\end{eqnarray*} 

It is important to emphasize that the first component of the objective function maximizes the violation considering the consumption of the other modes, and the second component, even for the single mode version, will add additional variables that do not contribute to the cut violation but which contribute to generate a stronger inequality.

\subsection{Lifted Precedence Based Cuts - LPR}
\label{subsec:prece}  

In addition to the cover cut introduced above, it is possible to further strengthen the formulation by analyzing the precedence between jobs and using precedence (PR) cuts similar to those used by \citet{Zhu2006}. Consider a job $j$, its (direct or indirect) successor $s$ and a time $t$. Also, consider the following constants $\displaystyle e_j=\textrm{min}_{m \in \mathcal{M}_j}(\mathcal{T}_{jm})$ and $\displaystyle l_j=\textrm{max}_{m \in \mathcal{M}_j}(\mathcal{T}_{jm})$ to limit the time period. We introduce a new lifted version for the precedence cuts, shown in Eq.(\ref{eq:ineqprec}). Consider the lengths of the shortest paths on the precedence graph, $\breve{d}_{jms}$ and $\breve{d}^{*}_{js}$, introduced in Section \ref{subsec:pre}.  The following inequalities are valid:
\begin{multline}
\label{eq:ineqprec}
\sum_{m \in \mathcal{M}_j} \sum_{t'= e_j}^{t+\breve{d}_{js}^{*}-\breve{d}_{jms}} x^*_{jmt'} \geq  
 \sum_{m \in \mathcal{M}_s} \sum_{t' = e_s}^{min(l_s, t+\breve{d}_{js}^{*})}  x^*_{smt'} \\ \forall j \in \mathcal{J}, s \in \overline{\mathcal{S}}_j , t \in \{ \max{(e_j, e_s - \breve{d}^*_{js} )} , \ldots , min(l_j,l_s -\breve{d}^*_{js})\}
\end{multline}

\paragraph{Example}  Consider the following two cuts. The first one, (PR), was generated with the RCPSP precedence cut as proposed in \cite{Zhu2006}. The second one, (LPR), is the lifted inequality (\ref{eq:ineqprec}) to predecessor job $5$ and its successor $10$ at time period $9$:
\begin{eqnarray*}
  (PR) = - x_{5,0,2} - x_{5,0,3} - x_{5,0,4} - x_{5,0,5} - x_{5,0,6} - x_{5,0,7} - x_{5,0,8} - x_{5,0,9} - \\ {\boxed{x_{5,0,10} - x_{5,0,11}}} - x_{5,1,2} - x_{5,1,3} - x_{5,1,4} - x_{5,1,5} - x_{5,1,6} - x_{5,1,7} - \\ x_{5,1,8} - {\boxed{x_{5,1,9} - x_{5,1,10} - x_{5,1,11}}} - x_{5,2,2} - x_{5,2,3} - x_{5,2,4} - x_{5,2,5} - \\ x_{5,2,6} -  {\boxed{x_{5,2,7} - x_{5,2,8} - x_{5,2,9}}} + x_{10,0,11} + x_{10,2,11} \leq 0 ;\\
  (LPR) = - x_{5,0,2} - x_{5,0,3} - x_{5,0,4} - x_{5,0,5} - x_{5,0,6} - x_{5,0,7} - x_{5,0,8} - x_{5,0,9} - \\ x_{5,1,2} - x_{5,1,3} - x_{5,1,4} - x_{5,1,5} - x_{5,1,6} - x_{5,1,7} - x_{5,1,8} - x_{5,2,2} -\\ x_{5,2,3} - x_{5,2,4} - x_{5,2,5} - x_{5,2,6} + x_{10,0,11} + x_{10,2,11} \leq 0.
\end{eqnarray*}

Notice the change in the coefficients of the variables on the left-hand-side of the lifted cut (LPR), based on Eq.(\ref{eq:ineqprec}) and corresponding to the variables of job $5$. In the case of different durations for processing modes of a job $j$, there will be an increase in the sum of the coefficients of the left-hand-side of (LPR) since we use $\breve{d}_{jms}$ instead of just considering the fastest mode given by $\breve{d}^{*}_{js}$ (as proposed in \cite{Zhu2006}). The original cut variant (PR) from \cite{Zhu2006} considers the fastest time. The variables highlighted in the boxes in the (PR) cut are not included in our cut approach (LPR). The lifted cut therefore strictly dominates the original cut, given that the highlighted variables have coefficients $0$ (instead of $-1$) in the cut stated as $\leq$ inequality.

Our separation procedure (see Algorithm \ref{alg:filtering}) therefore selects different paths (line \ref{alg:cp}) that connect job $j$ and the artificial project completion job  $a_{p}$. In particular, it is desirable to avoid redundant constraints, i.e., constraints of predecessor/successor jobs belonging to the same path in the dependency graph with a similar meaning, for jobs on subpaths that have already been used in previously added precedence cuts. Whenever a violated precedence cut is found on this path  (lines \ref{alg:cuts}--\ref{alg:endcuts}), the remaining jobs in this path are skipped (line \ref{alg:skipp}).  We limit the number $\zeta$ of precedence cuts that can be added per round and only add the most violated cuts  (line \ref{alg:zeta}).
 
\begin{algorithm}[!ht]\footnotesize
    \KwData{ set of jobs $\mathcal{J}$, artificial project completion job $a_{p}$, maximum number of cuts $\zeta$}
   \For{ (each $j \in \mathcal{J}$)} { \label{alg:J}
        $\breve{P} \gets$ \texttt{compute\_paths($j$,$a_p$)}; \label{alg:cp}\\ 
        $pos \gets 0$;\\
        \For{ (each $\breve{p} \in \breve{P}$)} { \label{alg:pth}
            $s \gets$ \texttt{successor($j$,$\breve{p}$,$pos$)};\\
            $found \gets 0;$\\
            \For{ ($ t \in \{ \max{(e_j, e_s - \bar{d}^*_{js} )} , \ldots , min(l_j,l_s -\bar{d}^*_{js})\}$)} {   \label{alg:periods}
                $ \mathbf{c} \gets \texttt{init\_cut}()$; \label{alg:cuts}\\ 
                 $(V,C) \gets $identifies a set of variables and their coefficients that generate a violated precedence cut for the predecessor $j$ and successor $s$ at time period $t$;\\
                 \If{ ($V  \neq \emptyset)$} {
                    $\texttt{add\_set\_var}(\mathbf{c}, V, C)$; \\  
                    $C \gets C \cup \{\texttt{cut}(\mathbf{c},\geq, 0 $ )\};\\
                    $found \gets 1;$  \label{alg:endcuts} \\
                }
            }
            \If{($found$)}{
                skipp the remaing jobs in the path of $j$ to $a_p$;\label{alg:skipp}\\
            }\Else{
                $pos \gets pos + 1$;
            }
        }
    }
    \texttt{sort\_cuts\_by\_highest\_violation\_and\_filters}($\mathcal{C}$, $\zeta$); \label{alg:zeta}\\
    \texttt{return $\mathcal{C}$;}
    \caption{\texttt{lifted\_precedence\_based\_cuts}}
    \label{alg:filtering}
\end{algorithm} 

\subsection{Conflict-Based Cuts: Cliques (CL) and Odd-Holes (OH)}
\label{subsec:conflict}

According to \citet{Padberg1973}, LP relaxations for problems that mostly contain binary variables linked to generalized upper bound (GUB) constraints can be significantly strengthened by the inclusion of inequalities derived from the set packing polytope (SPP). Generally, clique and odd-holes cuts can be generated using Conflict Graphs ($\mathcal{CG}$). The denser the $\mathcal{CG}$, the more inequalities can be generated. The disadvantage of having dense $\mathcal{CG}$ is that they can be prohibitively large \cite{atamturk2000}, so our algorithm creates the $\mathcal{CG}$ dynamically at each iteration by considering a set ($\mathcal{U}$) that contains the variables of interest, i.e., variables that have a non-zero value in the LP relaxation or variables set to zero but with a small reduced cost.

Several well-known inequalities can be generated considering pairwise conflicts between binary variables stored in $\mathcal{CG}$. Some conflicts can be easily detected by solvers considering constraints such as (\ref{model:maximoUmaVez}). Other conflicts can be implied from optimality conditions or by analyzing problem specific structures. Overall, the denser is the  $\mathcal{CG}$, stronger are the produced cuts.

The dynamic dense $\mathcal{CG}$ created is used in a separation procedure for inequalities derived from a common class of cuts for the SPP: cliques and odd-holes. A clique inequality for a set $C$ of conflicting variables has the form $\sum_{j \in C} x_j \leq 1$ and an odd-hole inequality for a cycle $C$ can be defined as: $\sum_{j \in C} x_j \leq \lfloor \frac{|C|}{2} \rfloor$ \cite{Santos2016}. 

\citet{Santos2016} present a clique separation routine that separates all violated cliques into a conflict subgraph induced by fractional variables. The authors then present a lifting that extends generated cliques considering the original $\mathcal{CG}$. They also present a strengthening of odd-holes inequalities by the inclusion of a so-called wheel center. For an odd-hole with variables $C$ and $W$ being the set of candidates to be included as wheel centers of $C$, the inequality (\ref{eq:ineqoh}) is valid: 

\begin{equation}
\sum_{j \in W} \lfloor \frac{|C|}{2} \rfloor x_j + \sum_{j \in C} x_j \leq  \lfloor \frac{|C|}{2} \rfloor 
\label{eq:ineqoh}
\end{equation}

Our approach presented in Algorithm \ref{alg:conflictgraph} considers four conflict types for variables $x_{jmt}$:

\begin{enumerate}
\item conflicts between variables of the same job (lines 6--7);
\item conflicts involving jobs that if allocated at the same time exceed the capacity of available renewable resources  (lines 8--10); 
\item conflicts based on precedence relations, in which the time window between the predecessor job $j$ on mode $m$ and some $s \in \overline{\mathcal{S}}_j$ is smaller than $\breve{d}_{jms}$  (lines 11--14);
\item conflicts considering jobs of different projects, where the sum of the delays generated by allocating these jobs in specific positions implies a total delay greater than $\alpha$ (lines 15--16).
\end{enumerate}

\begin{algorithm}[!ht]\footnotesize
    \KwData{ variables set $\mathcal{U}$, set $\mathcal{R}$, set  $\bar{\mathcal{S}}$,  delay $\alpha$ }
    \KwResult{Conflict Graph $\mathcal{C}\mathcal{G}$}
    $\mathcal{C}\mathcal{G} \gets \emptyset$;\\
    \For{ ($v_1 \in  \mathcal{U}$)} {  
         $j_1 \gets \texttt{job}(v_1)$;
         $m_1 \gets \texttt{mode}(v_1)$;
         $t_1 \gets \texttt{time}(v_1)$;\\
       	 \For{ ($v_2 \in \mathcal{U}$)} {	
	         $j_2 \gets \texttt{job}(v_2)$; 
    	     $m_2 \gets \texttt{mode}(v_2)$;
        	 $t_2 \gets \texttt{time}(v_2)$;\\

			\If{ ($j_1 = j_2$)}{
            	 $ \texttt{add\_var\_conf}(\mathcal{C}\mathcal{G},v_1,v_2)$; 
                 \textbf{continue};\\
             }
            
             \If{($t_1 = t_2$)}{
                 \For{ ($r \in  \mathcal{R} : ( q_{rj_1m_1} + q_{rj_2m_2} > \breve{q}_{r} )$)} { 
            	 		$ \texttt{add\_var\_conf}(\mathcal{C}\mathcal{G},v_1,v_2)$;
                 		\textbf{continue};\\
            	}
              }
  		 	 \If{($ j_2 \in \bar{\mathcal{S}}_{j_{1}}$)}{
               $w \gets \bar{d}^*_{j_1,j_2} - \texttt{job\_min\_dur}(j_1)$;\\
           	   \If{( ($\texttt{end\_time}(j_1)>t_2) \ or \ (t_2-\texttt{end\_time}(j_1)<w$))}{
            	 	$ \texttt{add\_var\_conf}(\mathcal{C}\mathcal{G},v_1,v_2)$; 
                    \textbf{continue};\\
        	 	}
         	}
            
             \If{($ {\texttt{proj}(j_1) \ != \  \texttt{proj}(j_2)\ \& } \  (\texttt{delay}(j_1,m_1,t_1) \ + \texttt{delay}(j_2,m_2,t_2)) > \alpha $)}{  	
            		 $ \texttt{add\_var\_conf}(\mathcal{C}\mathcal{G},v_1,v_2)$; 
             }
    	}
    }
	\textbf{return} $\mathcal{C}\mathcal{G}$;
    \caption{\texttt{creating\_conflict\_graph}}
\label{alg:conflictgraph}
\end{algorithm}

After the creation of the $\mathcal{CG}$, cuts can be generated considering the current fractional solution. In this paper, we use the routines described in \cite{Brito2015}, where cliques and odd-holes are exactly separated and lifted.

\paragraph{Example} Considering the following clique cut:
\begin{eqnarray*}
(CL) = x_{5,0,12} + x_{5,1,11} + x_{5,2,9} +\\ x_{10,0,11} + x_{10,0,12} + x_{10,0,13} + x_{10,2,11} + x_{10,2,12} + x_{10,2,13} \leq 1.
\end{eqnarray*}

It is possible to observe, from Figure \ref{fig:gantt}, that the variables corresponding to jobs $5$ and $10$ have conflicts at different time periods considering different modes, given that they have a precedence relation.

\paragraph{Example} Consider the following odd-hole cut: 
\begin{eqnarray*}
(OH) = x_{1,0,0} + x_{1,0,5} + x_{2,1,0} + x_{2,1,3} + x_{5,0,3} \leq 2.
\end{eqnarray*}

Still referring to job  $5$ in Figure \ref{fig:gantt}, we can observe conflicts from the precedence relationship with job $1$. Also, a low amount of resources available at times when jobs $5$ and $2$ intersect on the variables of the example above reflect conflicts between them. Thus, these three jobs can not be allocated in parallel, considering the amount of resources consumed by their modes.

\subsection{Strengthened Chv\'atal-Gomory Cuts - (SCG)}
\label{subsec:cg}

Chv\'atal-Gomory (CG) cuts are well-known cutting planes for MILP models. The inclusion of these cuts allows to significantly reduce the integrality gaps, even when only rank-one cuts are employed, i.e., those obtained from original problem constraints \cite{Fischetti2007}. 

Consider the integer linear programming (ILP) problem as min\{$\vec{c}^T \vec{x}: A\vec{x} \leq \vec{b}, \vec{x} \geq 0 $ integer\}, where  $A \in \mathbb{R}^{m x n}$, $ \vec{b} \in \mathbb{R}^m $, and $ \vec{c} \in \mathbb{R}^n $, with the two associated polyhedra $P := \{\vec{x} \in \mathbb{R}_{+}^{n}: A\vec{x} \leq \vec{b} \}$ and $P_{\mathcal{I}} := conv\{\vec{x} \in \mathbb{Z}_{+}^{n}: A\vec{x} \leq \vec{b} \} = conv (P\cap \mathbb{Z}^n) $ with $\vec{x}$ being integer variables. Consider $\mathcal{I}$ and $\mathcal{H}$ the sets of constraints and variables, respectively.

A Chv\'atal-Gomory cut \cite{Chvatal1973} is defined as a valid inequality for $P_{\mathcal{I}}$: $\lfloor \vec{u}^T A \rfloor \vec{x}$ $\leq \lfloor \vec{u}^T\vec{b} \rfloor$, where $\vec{u} \in \mathbb{R}_{+}^{m}$ is a multiplier vector.  The choice of $\vec{u} \in \mathbb{R}^{+}$ is crucial to deriving useful inequalities. Fischetti and Lodi \cite{Fischetti2007} propose the MILP model for Chv\'atal-Gomory separation. The maximally violated $\sum_{j \in \mathcal{H}(x^*)} {a_j  x_j} \leq a_0 $ inequality can be found by optimizing the following separation MILP model:
\\

\noindent \textit{Maximize:}
\begin{eqnarray}
\label{eq:cg1}
	\sum_{j \in \mathcal{H}(x^*)}{a_j  x^*_j} - a_0 
\end{eqnarray}
\noindent \textit{subject to:}
\begin{eqnarray}
\label{eq:cg2}
f_j = \vec{u}^T  A_j - a_j, \ \forall j \in \mathcal{H}(x*) \\ 
\label{eq:cg3}
f_0 = \vec{u}^T  \vec{b} - a_0 \\ 
\label{eq:cg4}
0 \leq f_j \leq 1 -  \delta \ \forall j \in \mathcal{H}(x*)\cup\{0\}\\
\label{eq:cg5}
-1+ \delta \leq u_i \leq 1 -  \delta \ \forall i = 1, \ldots, m \\
\label{eq:cg6}
a_j \ \in \ \mathbb{Z}, \ \forall j \in \mathcal{H}(x*) 
\end{eqnarray}
where $\mathcal{H}(x^*) := \{j \in {1, \ldots , \breve{n}} : x^*_j > 0\}$ and $x^*$ are fractional values for all $\breve{n}$ variables of an LP solution for a general problem fixed in the MILP. To strengthen the cut, a penalty term $\sum_{i}w_i  u_i$, with $w_i =10^{-4}$ for all $i$, is applied to the objective function. To improve the numerical accuracy of the method, multipliers too close to 1 are forbidden ($u_i \leq 0.99, \forall i$).

\paragraph{Example} Consider the following cut generated with the CG for jobs from Figure \ref{fig:gantt}: 
\begin{eqnarray*}
(CG) = 3 x_{1,0,8} + x_{5,0,5} + 2 x_{5,0,6} + 4 x_{6,1,4} + 2 x_{6,1,8} + 2 x_{8,1,7} + 2 x_{9,1,7} \leq 5.
\end{eqnarray*}

On the one hand, the larger the set of non-redundant and tight constraints considered in the Chv\'atal-Gomory separation the more likely it is that violated inequalities will be found. On the other hand, large separation problems can be hard to solve and the overall performance of the cutting plane method can degrade. The following subsection will therefore consider specific strategies to find suitable sets of constraints.

\subsubsection{Finding a Set of Constraints}

In our approach we consider a tuple ($\bar{s},\bar{f}$) to indicate the starting time and the finishing time of a given interval with size $\eta$ to filter the constraints and variables sets. We compute, for different intervals $(\bar{s},\bar{f})$, the summation of all infeasibilities for the integrality constraints for all their variables $x_{jmt}$ where $ \bar{s}  \leq t \leq \bar{f}$,  $x^*_{jmt}$ are the fractional values for the RCPSP variables. The value $\hat{f}$ is composed of the sum of the nearest integer distance, to indicate how fractional the variables in that interval are. 

\begin{equation}
    \hat{f}_{(\bar{s},\bar{f})} = \sum_{j \in \mathcal{J}}\sum_{m \in \mathcal{M}_j}\sum_{t=\bar{s}}^{\bar{f}}{|round(x_{jmt}^{*})-x_{jmt}^{*}|} \ \ \ \forall \ (\bar{s},\bar{f}) \in \mathcal{T}
\end{equation}

To find the most fractional interval $\hat{f}_{(\bar{s},\bar{f})}$ in the scheduling, in which the constraints with their respective variables will be chosen, we start from the beginning of the scheduling, and we slide the interval $(\bar{s},\bar{f})$ until the end of the scheduling (see Algorithm \ref{alg:constraints}, line \ref{alg:interval}). The parameters on the algorithm indicate the interval size $\eta$, the jump size $\iota$ to go to the next interval and a percentage $\zeta$ that allows sliding the interval. The slide only occurs if the violation is  greater than the current value, plus the percentage of $\zeta$. Another input data is the dynamic conflict graph $\mathcal{C}\mathcal{G}$.

\begin{algorithm}[!ht]\footnotesize
    \KwData{ fractional RCPSP solution  $x^*$, rows set $\mathcal{I}$, conflict graph $\mathcal{C}\mathcal{G}$, size interval  $\eta$, size jump $\iota$, percentage allowed to slide the interval $\zeta$}
    \KwResult{ Set  $\mathcal{Z}$ of selected rows}
    $\mathcal{V} \gets \emptyset$;\\    
    $(\bar{s},\bar{f}) \gets \texttt{identify\_interval}(x^*,\eta,\iota,\zeta)$;\label{alg:interval}\\
    \For{ ($(r_1 \in  \mathcal{I}) \ and \ \texttt{is\_in\_interval}(r_1, \bar{s},\bar{f})) $)} { \label{alg:resRInit} 
         $type \gets \texttt{type\_row}(r_1)$;\\
         \If{ ($type = Ineq.\ref{model:recursoR}$)}{
             $ \texttt{add\_row}(\mathcal{Z},r_1)$;\\
             $ \mathcal{V} \gets \mathcal{V} \cup \texttt{vars\_row}(r_1) $;\\
         }    
     }\label{alg:resREnd}
   $ O \gets \texttt{jobs}(\mathcal{V})$;  \label{alg:modetimeinit} \\
    \For{($ j \in O$)} {
         $\mathcal{V}_j \gets \texttt{vars\_jobs}(\mathcal{V},j)$;\\   
         $r_2 \gets \texttt{create\_constraints}(\leq,1)$;\\
         $ \texttt{add\_set\_var}(r_2,\mathcal{V}_j)$;
         $ \texttt{add\_row}(\mathcal{Z},r_2)$;\\
     }\label{alg:modetimeend}
     $ \mathcal{K} \gets \texttt{nonrenewable\_resources}(\mathcal{V})$; \label{alg:resRNInit} \\
     \For{($ k \in  \mathcal{K} $)}{ 
         	 $\mathcal{V}_k \gets \texttt{vars\_nonrenewable\_resources}(\mathcal{V},k)$;\\             
             $r_3 \gets \texttt{create\_constraints}(\leq,\texttt{capacity}(k))$;\\
             $ \texttt{add\_set\_var}(r_3,\mathcal{V}_k)$;
             $ \texttt{add\_row}(\mathcal{Z},r_3)$;\\
     }\label{alg:resRNend}
 	$ \mathcal{C} \gets \texttt{pairs\_of\_conflicting\_variables}(\mathcal{C}\mathcal{G},\mathcal{V})$; \label{alg:conflictinit}\\
    \For{($ c \in  \mathcal{C} $)}{ 
     	 $\mathcal{V}_c \gets \texttt{vars\_conflicts}(\mathcal{V},c)$;\\
         $r_4 \gets \texttt{create\_constraints}(\leq,1)$;\\
         $ \texttt{add\_set\_var}(r_4,\mathcal{V}_c)$;
         $ \texttt{add\_row}(\mathcal{Z},r_4)$;\\
     }\label{alg:conflictend}
	\textbf{return} $\mathcal{Z}$;
    \caption{\texttt{finding\_set\_constraints}}
\label{alg:constraints}
\end{algorithm} 

Preliminary experiments showed that cuts separated using constraints from the beginning of the time horizon were more effective for improving the dual bound, jobs allocated in the first time period are responsible for pushing the allocations of the others in the precedence graph. It is therefore desirable to find integer values for the first ones. Once the most fractional interval is found, the most important constraints are identified to compose the set for the Chv\'atal-Gomory separation.

The renewable resources directly impact the duration of the projects; therefore, all the variables of these constraints into the interval ($\bar{s},\bar{f}$) are considered in the set $\mathcal{V}$ (see Algorithm \ref{alg:constraints}, lines \ref{alg:resRInit}-\ref{alg:resREnd}). Further constraints from the original problem are included in the separation problem whenever they are related to the current time window: constraints that restrict the choice of only job (lines \ref{alg:modetimeinit}-\ref{alg:modetimeend}) and non-renewable resource constraints (lines \ref{alg:resRNInit}-\ref{alg:resRNend}). Functions \texttt{nonrenewable\_resources}($\mathcal{V}$) and \texttt{jobs}($\mathcal{V}$) returns, respectively, the set of non-renewable resources and jobs where variables of $\mathcal{V}$ appear. A good strategy is to find additional constraints that represent conflicts between the variables of set $\mathcal{V}$. A main conflict is analyzed, whereby the time window comprising variables coming from the conflict graph $\mathcal{C}\mathcal{G}$ generated dynamically at each iteration of the cutting plane received as input parameter (lines \ref{alg:conflictinit}-\ref{alg:conflictend}). Function \texttt{pairs\_of\_conflicting\_variables}($\mathcal{C}\mathcal{G},\mathcal{V}$) returns the pairs of these conflicting variables.

\subsubsection{Strengthening Procedure}
\label{subsubsec:strengthening}

To produce strengthened CG cuts, a strategy similar to the proposal of \citet{Letchford2016} is employed. The key idea is to take violated CG cuts and then strengthen the right-hand-sides ($rhs$). \citet{Letchford2016} solves the maximum weight stable set problem for the conflict graph induced by the binary variables of the CG cut to find a (hopefully smaller) new valid $rhs$. Our approach solves the same problem augmented by additional constraints involving these variables. Consider here that set $\mathcal{H}$ contains all variables of the cut to be strengthened. $A_{\mathcal{H}}$ is the matrix of coefficients of $\mathcal{H}$ including non-renewable and renewable resource constraints, job allocation constraints and conflict constraints, with $rhs$ values specified in a vector $\vec{b}$. 
The vector $\vec{c}$ are the coefficients of the variables that appear in the cut. Consider vector $\vec{x}$ as variables of $\mathcal{H}$ and the integer linear programming as max\{$\vec{c}^{T}\vec{x}: A_{\mathcal{H}}\vec{x}  \leq \vec{b}$\}. If the optimal solution value of this MILP is smaller than the original $rhs$ of the cut, the original CG can be strengthened with this new value on the $rhs$.

\paragraph{Example} Considering the previous example, it is possible to strengthen the Chv\'atal-Gomory cut by tightening the value on the right side to $4$ based on the MILP presented before. Notice that by applying the strengthening, it was possible to reduce the rhs value by 20\%:
\begin{eqnarray*}
(SCG) = 3 x_{1,0,8} + x_{5,0,5} + 2 x_{5,0,6} + 4 x_{6,1,4} + 2 x_{6,1,8} + 2 x_{8,1,7} + 2 x_{9,1,7} \leq \boldsymbol{4}.
\end{eqnarray*}

\section{Computational Results}
\label{sec:results}

This section presents the results of the experiments obtained by the proposed cutting plane algorithm and the preprocessing routine to strengthen resource-related constraints. All computational experiments have been carried out on a computing cluster (Compute Canada) composed by Intel \textregistered\,  Xeon X5650 Westmere processors with 2,67 GHz and 512 GB of RAM running Scientific Linux release 6.3. All algorithms were coded in \texttt{ANSI C 99} and compiled with GCC version 5.4.0, with flags \emph{-Ofast} and solver GUROBI version 8.0.1 \cite{gurobi}.

\subsection{Benchmark Datasets}
\label{sec:inst}
In order to facilitate the comparison of algorithmic improvements, several public databases were established, including realistic, but computationally challenging problem instances. In 1996 a public and well-known repository of benchmark datasets, devoted to this class of problems, was established: the Project Scheduling Problem Library - PSPLIB \cite{Kolisch1996}. In 2013 a new dataset based on PSPLIB instances, considering multiple projects, emerged from the MISTA Challenge \cite{Wauters2016}. Then, in 2014, another dataset based on PSPLIB was proposed for the multi-mode scheduling problem called MMLIB \citep{VanPeteghem2014}. For all problem variants, many of those instances are still open. In particular, by the time of writing this paper, we found that 215 of the PSPLIB instances, 2842 of the MMLIB instances, and 27 of the MISTA instances, were still open\footnote{1339 considering the PSPLIB website and 30 considering the MISTA website. There are no informations about the optimal solutions achieved by \cite{Schnell2017} and \cite{Toffolo2016}.}.

We now define the benchmark used in this work based on \cite{Schnell2017,Toffolo2016,Zhu2006} and on instances from the PSPLIB, MISTA and MMLIB repositories. From PSBLIB, we consider instances for the SMRCPSP and MMRCPSP problem variants. From MISTA, we included all problem instances from the basic dataset A for the MMRCMPSP. For MMRCPSP, we also use all instances from MMLIB in the final experiments. 

Table \ref{tab:benchmarkdataset} shows, for each library, the total number of problem instances defined for each problem variant, the group name, the number of instances that are used in our benchmark datasets, and the number of instances for which optimality has not yet been proven in the literature.

\setlength\tabcolsep{2pt}
\begin{table}[!ht]
    \footnotesize
    \centering
     \caption{Benckmark datasets}
    \begin{tabular}{r|r|r|r|r}
     \hline
       \multicolumn{1}{c|}{library} &
       \multicolumn{1}{c|}{variant/num. inst.} & \multicolumn{1}{c|}{group} & \multicolumn{1}{c|}{in dataset} & \multicolumn{1}{c}{open}  \\ \hline
\multirow{5}{*}{PSPLIB}  & \multirow{3}{*}{SMRCPSP/2040} & J60 & 79 & 79 \\
          &  & J90 & 105  & 105  \\
          &  & J120 &  514 &  514 \\ \cline{2-5}
          &  \multirow{1}{*}{MMRCPSP/3931}  & J30  &  641 & 31  \\ \hline
MISTA   & MMRCMPSP/30 & A & 10  & 7 	\\ \hline
\multirow{4}{*}{MMLIB}	& \multirow{4}{*}{MMRCPSP/4300}  & J50 &  540 &  118 \\
	&   & J100 & 540 & 176  \\
	&   & J50+ & 1620 &  1178 \\
	&   & J100+ & 1620 & 1370  \\\hline
    \end{tabular}
    \label{tab:benchmarkdataset} 
\end{table}

\subsection{Cutting Plane Algorithm Experiments}
\label{sec:cpexp}

In order to evaluate the performance of the proposed cutting plane in relation to the specific problem cuts and the linear relaxation, preliminary experiments have been performed on the benchmark datasets with $\alpha > 0$\footnote{$\alpha > 0$ is an upper  bound  to  the maximum TPD with value higher then its CPD.} totaling 782 instances from PSPLIB and MISTA. These two sets include instances of the versions SMRCPSP, MMRCPSP and MMRCMPSP. Instances from MMLIB, that contemplate a larger number of varied instances for the MMRCPSP version, were introduced only in experiments carried out using the complete version of our approach. The impact of adding the cuts for instances where  $\alpha = 0$ could not be measured, since the LP bound remains at the same value even when the formulation has been improved. We used a time limit of 24 hours for each instance from the MMRCMPSP and 4 hours for each instance from the SMRCPSP and MMRCPSP.

The first experiment was conducted to evaluate the preprocessing MILP formulation by analyzing the LP relaxation (LR) and the strengthened LP relaxation (SLR) using the coefficient strengthening MILP presented in Section \ref{subsec:premip} to renewable resources constraints. The reported computing time for the SLR includes the time spent to find the successful feasible sets with parameters to stop the enumeration process as $it=200,000$ and $tl$ as the maximum allowed time defined above to run the approach. Table \ref{tab:lrlrs} shows the average integrality gaps \footnote{given the best known upper bound $\overline{b}$, from PSPLIB and MISTA websites, and an obtained dual bound $\underline{b}$, the integrality gap is computed as follows: $\frac{(\overline{b}-\underline{b})}{\overline{b}}$, $\overline{b}>0$} and the average computing times in seconds that have been obtained with the original LP relaxations and the strengthened LR for the instances from PSPLIB and MISTA.

\setlength\tabcolsep{2pt}
\begin{table}[!ht]
    \footnotesize
    \centering
     \caption{Average integrality gaps and the average computing times (sec.) for solving with the original LP relaxations (LR) and the strengthened LR (SLR)}
    \begin{tabular}{rr|rr|rr}
     \hline
     & & \multicolumn{2}{c|}{LR} & \multicolumn{2}{c}{SLR}\\
     \multicolumn{1}{c}{group}	& \multicolumn{1}{c|}{$n$} & \multicolumn{1}{c}{gap}	&	\multicolumn{1}{c|}{time} &  \multicolumn{1}{c}{gap}	& \multicolumn{1}{c}{time}  \\ \hline
A	    & 10  & 0.441 & 85.9 & \textbf{0.438} & 683.6 \\
J30	    & 245 & 0.663 & 0.3 & \textbf{0.658} & 2.8  \\
J60	    & 57 & 0.825 & 1.0 & \textbf{0.816} & 5.3  	\\
J90	    & 80 & 0.822 & 2.1 & \textbf{0.818} & 26.9	 \\ 
J120	& 390 & 0.840 & 3.3 & \textbf{0.835} & 37.2  \\ \hline
total	& 782 & 0.718 & 18.5 & \textbf{0.713} & 151.2  \\ \hline
    \end{tabular}
    \label{tab:lrlrs} 
\end{table}

The results in Table \ref{tab:lrlrs}, shown in bold numbers, indicate that the strengthening formulation slightly improves the integrality gaps. 
As expected, solving only the weak initial formulation is faster than solving the formulation with the preprocessing routines. However, we note that even small improvements in the root node can result in a large number of nodes pruned later in the search tree. For this reason, and given that the additional computing times are still quite reasonable, we use this strengthening strategy in all further experiments.

According to \citet{Kolisch1995}, network precedence relationships and the factor between availability-consumption of resources are the two critical characteristics of the instances. When pruning is applied considering these two characteristics, we note that even for instances with $120$ jobs, there are time periods for which it is possible to enumerate the feasible subsets. For example, it is possible to enumerate until $t=102$ for the large instance $j1201\_1$. This instance is composed of $120$ jobs and has restricted relationship between the availability and consumption of renewable resources.

\subsubsection{Results for Different Cut Families}

In order to devise an effective cutting plane strategy, we next evaluate the different separation strategies. In Section \ref{sec:cpexp}, it has been shown that using the SLR instead of the original LR strengthens the formulation while only marginally increasing computing times. We now explore the bound improvement when combining the SLR with each of the different cut types. Tables \ref{tab:results} and \ref{tab:resultsA} show the average integrality gaps and the average computing times for different approaches. In both tables, the first column SLR presents results obtained by the strengthened LP relaxation without cuts. The LCV\footnote{$\omega=100000$ and $\mu=0.1$}, LPR, CL, OH ad SCG columns indicate, respectively, results\footnote{the maximum number of precedence and clique cuts added to the LP at each iteration corresponds to 20\% of the amount of the LP rows} obtained by combining the SLR with one additional cut type: lift operations to cover and precedence cuts, clique cuts, odd-hole cuts and strengthened Chv\'atal-Gomory cuts. 

\setlength\tabcolsep{1.5pt}
\begin{table*}[!ht]
    \footnotesize
    \centering
     \caption{Average integrality gaps and average computing times (sec.) obtained by different cuts for the SMRCPSP and MMRCPSP benckmark datasets with the time limit of $4$ hours}
    \begin{tabular}{rr|rr|rr|rr|rr|rr|rr}
     \hline
     & & \multicolumn{2}{c|}{SLR} & \multicolumn{2}{c|}{+\{LCV\}} &  \multicolumn{2}{c|}{+\{LPR\}} &   \multicolumn{2}{c|}{+\{CL\}} & \multicolumn{2}{c|}{+\{OH\}} 	&	\multicolumn{2}{c}{+\{SCG\}}\\
     \multicolumn{1}{c}{group}	& \multicolumn{1}{c|}{$n$} & \multicolumn{1}{c}{gap}	&	\multicolumn{1}{c|}{time} 	&  \multicolumn{1}{c}{gap}	&	\multicolumn{1}{c|}{time} &  \multicolumn{1}{c}{gap}	&	\multicolumn{1}{c|}{time}	& \multicolumn{1}{c}{gap}	&	\multicolumn{1}{c|}{time} &	\multicolumn{1}{c}{gap}	&	\multicolumn{1}{c|}{time} &	\multicolumn{1}{c}{gap}	&	\multicolumn{1}{c}{time}  \\ \hline
J30	    & 245 & 0.658  & 3 & 0.644 & 6  & \textbf{0.549} & 5 & 0.567	& 101	& 0.657	& 3   & 0.551	& 14200	\\
J60	    & 57 & 0.816 & 5 & 0.814 & 11 & \textbf{0.712} & 88 & 0.806	& 1093 & 0.816	& 12	& 0.812	& 14401	\\
J90	    & 80 & 0.818 & 27	& 0.816 & 45 & \textbf{0.672} & 454	& 0.807	& 3251 & 0.818	& 35	& 0.815	& 14400	\\
J120	& 390 & 0.835 &	37 & 0.820 & 114 & \textbf{0.714} & 2111 & 0.817 & 7265 & 0.835 & 56 & 0.832 & 14423	\\  \hline
total	& 772 & 0.781 & 18.1 & 0.774 & 43.9 & \textbf{0.661} & 665 & 0.749	& 2928 & 0.781	& 27	& 0.780	& 14408	\\ \hline
    \end{tabular}
    \label{tab:results} 
\end{table*}

The results in Table \ref{tab:results} suggest that the best average values (in bold numbers) of the integrality gaps were obtained by adding the lifted precedence cuts. Those cuts offered the best bound improvement (about 15\%), while requiring only moderate additional computing time. For the complete cutting plane, in the final experiment, we execute the separation procedures in parallel, but if a hierarchical implementation approach was used, one would add the cut families based the order of their integrality gap improvements: LPR, CL, LCV, SCG and OH.

The results of instances from MMRCMPSP are presented in Table \ref{tab:resultsA}. The results summarized indicate that the LPR cuts are also effective for the multi-project problem variant, even for large instances. SCG cuts have been found to be particularly useful for this class of problem. Experiments for instances as from A-7 typically exceeded the given time or memory limits to insertion of cuts such as CL, OH, and SCG.

\setlength\tabcolsep{1.5pt}
\begin{table*}[!ht]
    \footnotesize
    \centering
    \caption{Integrality gaps and computing times (sec.) obtained by different cuts to the MMRCMPSP benckmark datasets with the time limit of $24$ hours}
   \begin{tabular}{r|rr|rr|rr|rr|rr|rr}
     \hline
      &  \multicolumn{2}{c|}{SLR} & \multicolumn{2}{c|}{+\{LCV\}} &  \multicolumn{2}{c|}{+\{LPR\}} &   \multicolumn{2}{c|}{+\{CL\}} & \multicolumn{2}{c|}{+\{OH\}} 	&	\multicolumn{2}{c}{+\{SCG\}}\\
     inst. & \multicolumn{1}{c}{gap}&	\multicolumn{1}{c|}{time} 	&  \multicolumn{1}{c}{gap}	&	\multicolumn{1}{c|}{time} &  \multicolumn{1}{c}{gap}	&	\multicolumn{1}{c|}{time}	& \multicolumn{1}{c}{gap}	&	\multicolumn{1}{c|}{time} &	\multicolumn{1}{c}{gap}	&	\multicolumn{1}{c|}{time} &	\multicolumn{1}{c}{gap}	&	\multicolumn{1}{c}{time}  \\ \hline
A-1	&	1.000	&	0	&	1.000	&	0	&	0.875	&	0	&	0.875	&	0	&	1.000	&	0	&	\textbf{0.000}	&	1793	\\
A-2	&	1.000	&	1	&	1.000	&	4	&	0.987	&	4	&	0.989	&	2	&	1.000	&	2	&	\textbf{0.935}	&	86007	\\
A-3	&	\textbf{0.000}	&	8	&	\textbf{0.000}	&	9	&	\textbf{0.000}	&	15	&	\textbf{0.000}	&	15	&	\textbf{0.000}	&	15	&	\textbf{0.000}	&	62097	\\
A-4	&	0.414	&	4	&	0.411	&	35	&	\textbf{0.333}	&	39	&	0.366	&	1297	&	0.414	&	10	&	0.396	&	86005	\\
A-5	&	0.305	&	32	&	0.299	&	1743	&	\textbf{0.273}	&	1168	&	0.301	&	6157	&	0.305	&	254	&	0.302	&	86028	\\
A-6	&	0.428	&	596	&	0.427	&	4118	&	\textbf{0.317}	&	3303 	&	0.413	&	34106	&	0.428	&	1724	&	0.424	&	86005	\\ \hline
total	&	0.5245	&	221	&	0.523	&	985	&	\textbf{0.464}	&	755	&	0,491	&	6930	& 0.525	&	334	&	\textbf{	0.343}	&	40983		\\ \hline
    \end{tabular}
    \label{tab:resultsA} 
\end{table*}

In Table \ref{tab:lift} we evaluate the lifting strategies of the traditional RCPSP cuts proposed by \cite{Zhu2006} and the strengthening strategy for Chv\'atal-Gomory cuts, also combining with the SLR.
\setlength\tabcolsep{1.5pt}
\begin{table*}[!ht]
    \footnotesize
    \centering
     \caption{Average integrality gaps and average computing times (sec.) obtained by lifting the traditional RCPSP cuts, by strengthening the CG cuts and their original versions combining with the SLR with time limit of $24$ hours for A group and $4$ hours for the others}
    \begin{tabular}{rr|rr|rr|rr|rr|rr|rr}
     \hline
     & & \multicolumn{2}{c|}{+\{CV\}} & \multicolumn{2}{c|}{+\{LCV\}}
     & \multicolumn{2}{c|}{+\{PR\}} & \multicolumn{2}{c|}{+\{LPR\}}
     & \multicolumn{2}{c|}{+\{CG\}} & \multicolumn{2}{c}{+\{SCG\}} \\
     \multicolumn{1}{c}{group}	& \multicolumn{1}{c|}{$n$} & \multicolumn{1}{c}{gap}	&	\multicolumn{1}{c|}{time} &  \multicolumn{1}{c}{gap}	& \multicolumn{1}{c|}{time}
     &  \multicolumn{1}{c}{gap}	& \multicolumn{1}{c|}{time}
     &  \multicolumn{1}{c}{gap}	& \multicolumn{1}{c|}{time}
     &  \multicolumn{1}{c}{gap}	& \multicolumn{1}{c|}{time}
     &  \multicolumn{1}{c}{gap}	& \multicolumn{1}{c}{time}
     \\ \hline
A    & 6  & 0.526	&	59.9	&	\textbf{0.523}	&	984.8 & 0.518 	& 555.9  &	\textbf{0.464}	& 754.8	& \textbf{0.342} & 71825 & 0.343 & 67989	\\
J30  & 245 & 0.660	&	0.7	    &	\textbf{0.644}	&	5.6	    &	0.653	&	3.7	    &	\textbf{0.549}	&	4.9    & \textbf{0.548} & 14092 & 0.551 & 14200 \\
J60  & 57  & 0.825	&	2.1	    &	\textbf{0.814}	&	11.2	&	0.794	&	30.1	&	\textbf{0.712}	&	87.5   & \textbf{0.811} & 14402 & 0.812 & 14401\\
J90  & 80  & 0.821	&	4.3	    &	\textbf{0.816}	&	44.9	&	0.780	&	131.8	&	\textbf{0.672}	&	454.3  & \textbf{0.815} & 14402 & \textbf{0.815} & 14400	 	\\ 
J120 & 390 & 0.839	&	66.2	&	\textbf{0.820}	&	113.9	&	0.809	&	1516.2	&	\textbf{0.714}	&	2111.4 & \textbf{0.832} & 14402 & \textbf{0.832} & 14423 \\
\hline
total & 782 & 0.734	&	26.6	&	\textbf{0,723}	& 232.1	& 0.711	& 447.5	&	\textbf{0.622}	& 682.6 & \textbf{0.670} & 25825 & 0.671 &  25083	\\ \hline
    \end{tabular}
    \label{tab:lift} 
\end{table*}

The lifting strategies were able to improve the average integrality gaps of all benchmark datasets when comparing with the traditional RCPSP cuts (CV and PR). The results in bold numbers suggest that the lifting strategy is generally successful, substantially improving the average integrality gaps while increasing computing times. Note that even for the benckmark datasets of SMRCPSP, improved lower bounds were achieved, since lifting tends to use more variables in the cut apart from those that contribute to the cover violation. The strengthened CG cuts did not improve upon the original CG cuts, which can be explained by the large computing times spent by the strengthening procedure. When analyzing the average number of iterations for both procedures, the CG without lift did 3118 iterations in the average, while SCG did just 2450 iterations. Even with a reduced number (21\%) of iterations it was able to achieve basically the same results.
 
\begin{figure}[!ht]
	\begin{center} \includegraphics[width=.85\linewidth]{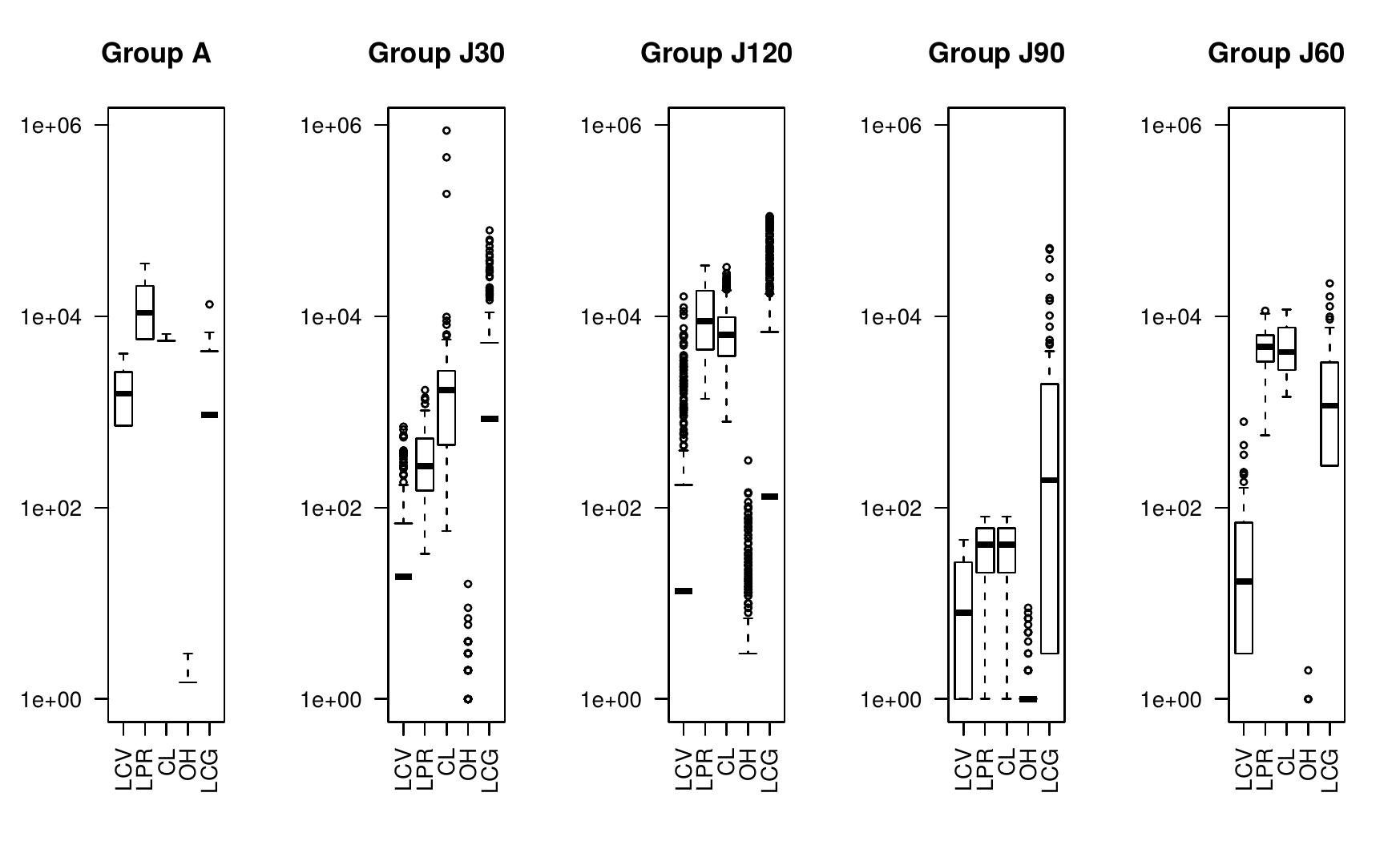}  
               \caption{Boxplots of the number of cuts for different types and benchmark datasets to each separation strategy}
             \label{fig:Cuts}
 \end{center}
\end{figure}

To analyze the cut generation for each separation strategy, we have computed the number of unique cuts for each strategy to the previous experiment. Figure \ref{fig:Cuts} shows boxplots of the number of cuts for different instance types and benchmarks.

 The lifted precedence strategy LPR finds a reasonable number of cuts for almost all benchmark datasets. The strengthened Chv\'atal-Gomory SCG strategy seems to find cuts for all the benchmark datasets, including J30 and J90 in which LPR finds less. Another peculiarity of SCG is that although it finds few cuts at each iteration it remains to find cuts for a larger number of iterations. It is also possible to observe that a small number of odd-holes cuts are generated, which explains the little impact on the integrality gaps. Finally, the quantities of both cliques and lifted precedence cuts are relatively high. However, their separation requires more time.

\subsubsection{Results Removing Cut Families}

While the previous experiments explored the impact of adding each cut type, we now explore the impact of omitting each cut type. To this end, we use all cut types, separated in parallel in the cutting plane, and then individually remove each type for the benchmark datasets SMRCPSP and MMRCPSP. Table \ref{tab:resultscp} summarizes the results of these experiments, where column pair All Cuts represents the strategy using all cut families, and the other column pairs represent the strategy without each of the cut types.

\setlength\tabcolsep{1.5pt}
\begin{table*}[!ht]
    \footnotesize
    \centering
    \caption{Average integrality gaps and the average computing times (sec.) obtained by the cutting plane using all cut types and then removing one cut type at a time with time limit of $4$ hours}
    \begin{tabular}{rr|rr|rr|rr|rr|rr|rr}
    \hline
 & & \multicolumn{2}{c|}{All Cuts} & \multicolumn{2}{c|}{-\{LCV\}} &   \multicolumn{2}{c|}{-\{LPR\}} & \multicolumn{2}{c|}{-\{CL\}} 	&	\multicolumn{2}{c|}{-\{OH\}} 	&	\multicolumn{2}{c}{-\{SCG\}}\\
     \multicolumn{1}{c}{group}	& \multicolumn{1}{c|}{$n$} & \multicolumn{1}{c}{gap}	& \multicolumn{1}{c|}{time}	&	\multicolumn{1}{c}{gap} &  \multicolumn{1}{c|}{time}	&	\multicolumn{1}{c}{gap}	& \multicolumn{1}{c|}{time}	&	\multicolumn{1}{c}{gap} &	\multicolumn{1}{c|}{time}	&	\multicolumn{1}{c}{gap} &	\multicolumn{1}{c|}{time} &
     \multicolumn{1}{c}{gap} &	\multicolumn{1}{c}{time} \\ 
     \hline
J30  	&	 245 	&	 \textbf{0.458} 	&	13546	&	0.471	&	13631	&	0.485	&	13640	&	0.459	&	13645	&	0.461	&	13661	&	0.518	&	8	\\
J60  	&	 57 	&	 \textbf{0.702} 	&	14402	&	0.703	&	14401	&	0.798	&	14401	&	 \textbf{0.702} 	&	14401	&	 \textbf{0.702} 	&	14402	&	0.71	&	107	\\
J90  	&	 80 	&	0.666	&	14403	&	0.667	&	14403	&	0.8	&	14402	&	 \textbf{0.665} 	&	14403	&	0.666	&	14403	&	0.669	&	545	\\ 
J120 	&	 390 	&	0.668	&	14409	&	0.713	&	14436	&	0.774	&	14425	&	 \textbf{0.667} 	&	14429	&	0.67	&	14429	&	 \textbf{0.667} 	&	4779	\\ \hline
Total 	&	 772 	&	0.624	&	14190	&	0.639	&	14218	&	0.714	&	14217	&	 \textbf{0.623} 	&	14219	&	0.625	&	14224	&	0.641	&	1360	\\ \hline
    \end{tabular}
    \label{tab:resultscp} 
\end{table*}

By removing the LPR cuts, results worsen by about 13\%. The results also get a little worse by removing OH, LCV, and SCG respectively. 
The results using all cuts performed generally well. In addition, removing the clique cuts further improved the gaps on larger instances. This can be explained by the fact that separating cliques requires more time. When these cuts are removed the numbers of iterations increases, as one can observe in Table \ref{tab:resultscpround}. Based on these results, one may define a new hierarchical sequence of the cut types based on the gap increase when the cut type is removed: LPR, SCG, LCV, OH, and CL.
\setlength\tabcolsep{1.5pt}
\begin{table*}[!ht]
    \footnotesize
    \centering
    \caption{Average number of iterations and the average computing times (sec.) obtained by the cutting plane comparing with removing some others cuts with time limit of $4$ hours}
    \begin{tabular}{rr|rr|rr|rr|rr|rr|rr}
    \hline
 & & \multicolumn{2}{c|}{All Cuts} & \multicolumn{2}{c|}{-\{LCV\}} &   \multicolumn{2}{c|}{-\{LPR\}} & \multicolumn{2}{c|}{-\{CL\}} 	&	\multicolumn{2}{c|}{-\{OH\}} 	&	\multicolumn{2}{c}{-\{SCG\}}\\
     \multicolumn{1}{c}{group}	& \multicolumn{1}{c|}{$n$} & \multicolumn{1}{c}{round}	& \multicolumn{1}{c|}{time}	&	\multicolumn{1}{c}{round} &  \multicolumn{1}{c|}{time}	&	\multicolumn{1}{c}{round}	& \multicolumn{1}{c|}{time}	&	\multicolumn{1}{c}{round} &	\multicolumn{1}{c|}{time}	&	\multicolumn{1}{c}{round} &	\multicolumn{1}{c|}{time} &
     \multicolumn{1}{c}{round} &	\multicolumn{1}{c}{time} \\ 
     \hline
J30  	&	 245 	&	2077	&	13546	&	1946	&	13631	&	1796	&	13640	&	2486	&	13645	&	2035	&	13661	&	12	&	8	\\
J60  	&	 57 	&	618	&	14402	&	618	&	14401	&	877	&	14401	&	832	&	14401	&	452	&	14402	&	16	&	107	\\
J90  	&	 80 	&	294	&	14403	&	300	&	14403	&	476	&	14402	&	401	&	14403	&	293	&	14403	&	22	&	545	\\ 
J120 	&	 390 	&	265	&	14409	&	271	&	14436	&	347	&	14425	&	469	&	14429	&	284	&	14429	&	170	&	4779	\\ \hline
Total 	&	 772 	&	814	&	14190	&	784	&	14218	&	874	&	14217	&	1047	&	14219	&	766	&	14224	&	55	&	1360	\\ \hline
    \end{tabular}
    \label{tab:resultscpround} 
\end{table*}

Table \ref{tab:resultscpA} shows the results for the same experiments for the MMRCMPSP. Again, column All Cuts represent the strategy were all cut types are used. Instead of reporting the results when each of the other cut types is removed, we only report on removing the cut types that consumed most of the computing time: cliques and odd-holes cuts, or the strengthened CG cuts. The bold numbers show the best average integrality gaps achieved for instances that did not run out of memory. 
Even though the results are better on average when using all cuts, by removing the cuts, the time was reduced significantly.  Even though using all cut types significantly increases the computing times, the gap improvement on some of the instances may still justify their use. 

 \setlength\tabcolsep{1.5pt}
\begin{table}[!ht]
    \footnotesize
    \centering
    \caption{Average integrality gaps and the average computing times (sec.) obtained by the cutting plane comparing with removing some others cuts with time limit of $24$ hours}
    \begin{tabular}{r|rr|rr|rr}
    \hline
 &  \multicolumn{2}{c|}{All Cuts} &  \multicolumn{2}{c|}{-\{CL\&OH\}} & \multicolumn{2}{c}{-\{SCG\}} \\
      \multicolumn{1}{c|}{instance} &
      \multicolumn{1}{c}{gap}	&	\multicolumn{1}{c|}{time} & \multicolumn{1}{c}{gap}	&	\multicolumn{1}{c|}{time} &  \multicolumn{1}{c}{gap}	&	\multicolumn{1}{c}{time}	 \\ 
     \hline
A-1 	&	\textbf{ 0.000} 	&	    52.5 	&	0.875	&	  0.7 	&	 0.875 	&	    0.4 	\\	
A-2 	&	 \textbf{0.781} 	&	 86015.2 	&  0.924	&	  4.2 	&	 0.924 	&	    5.6 	\\	
A-3 	&	\textbf{ 0.000} 	&	 86368.5 	&	 \textbf{0.000} 	&	  11.0 	&	\textbf{ 0.000} 	&	    9.4 	\\	
A-4 	&	 \textbf{0.316} 	&	 86001.1 	&	0.327	&	 31.1 	&	 0.327 	&	  104.2 	\\	
A-5 	&	 \textbf{0.263} 	&	 86010.4 	&	0.267	&	 832.2 	&	 0.266  	&	 1280.4 	\\	
A-6 	&	 \textbf{0.313} 	&	 86089.2 	&	0.314	&	 2130.2 	&	 0.314 	&	 2660.6 	\\	\hline
Total	&	 \textbf{0.279} 	&	 71756.2 	&	0.451	&	501,6	&	 0.451  	&	 676.8 	\\	\hline
    \end{tabular}
    \label{tab:resultscpA} 
\end{table}

\begin{figure}[!htbp]
	\begin{center} \includegraphics[width=.85\linewidth]{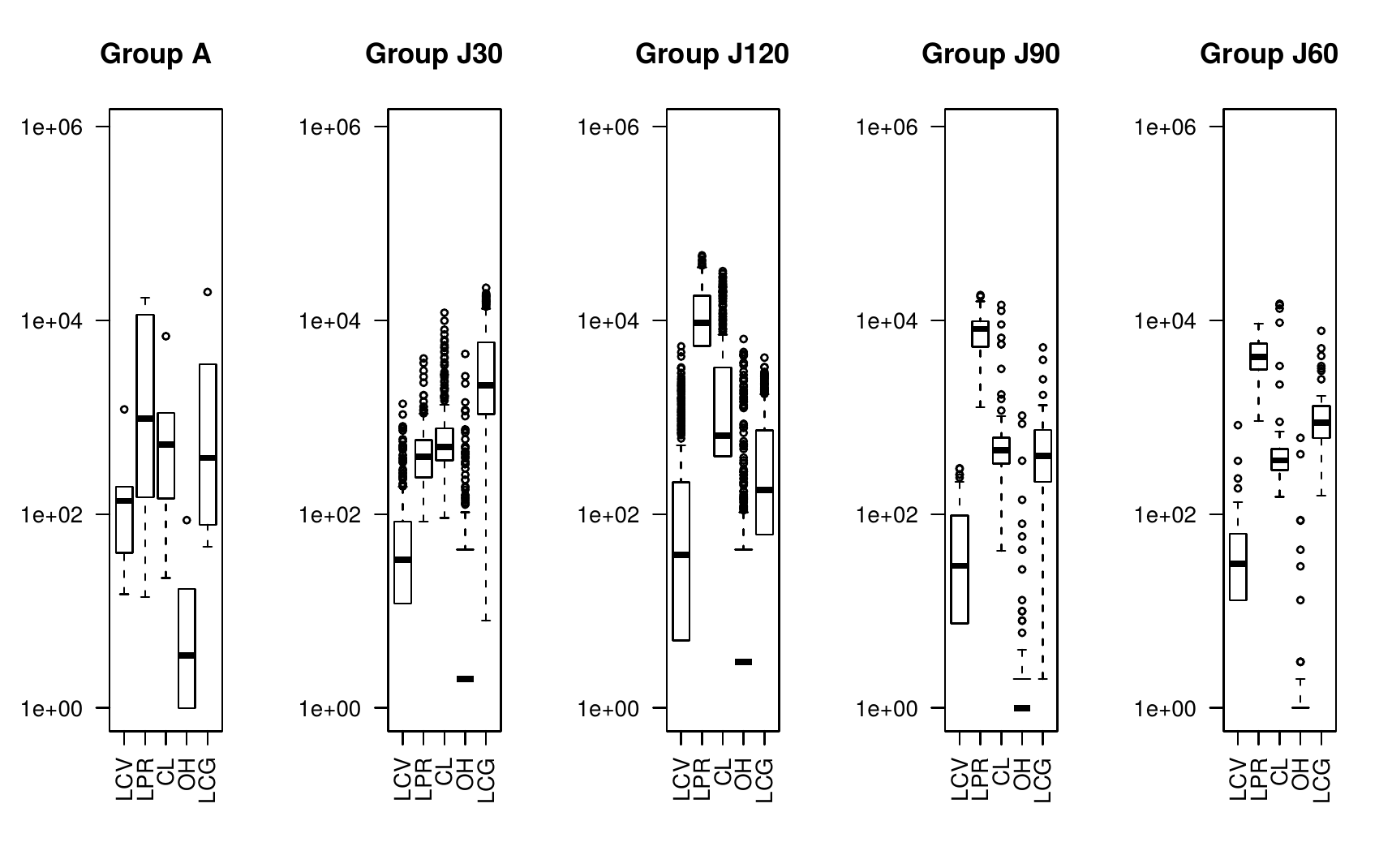}
               \caption{Boxplots of the number of cuts for different types and benchmark datasets to all separation strategy together}
             \label{fig:AllCuts}
 \end{center}
\end{figure}

To analyze the number of generated cuts for all separation procedures, we also computed the number of non-repeated cuts to the previous experiments (All Cuts). Figure \ref{fig:AllCuts} shows the boxplots representing the number of cuts for different types and benchmarks.

Figure \ref{fig:AllCuts} shows that LPR cuts are effective for all datasets mainly in the case of the SMRCPSP, where it found, on average, more cuts than others. For the MMRCMPSP, the LPR cuts are still the most effective ones. CL and SCG cuts seem to be effective mainly for instances with multiple modes. The average number of generated LCV cuts seems to be stable for all benchmarck dataset. Few OH cuts are found for all benchmarck datasets.

After analyzing the best strategy for the MMRCPSP obtained through experiments on the PSBLIB instances, we run the experiment with all cuts for the instances from MMLIB, comparing the LP relaxation and the strengthened linear relaxation. The results are presented in Table \ref{tab:resultscpMMLIB}. 

\setlength\tabcolsep{1.5pt}
\begin{table*}[!ht]
    \footnotesize
    \centering
    \caption{Average integrality gaps and the average computing times (sec.) obtained by the cutting plane comparing with the LP relaxation and its strengthening with time limit of $4$ hours}
    \begin{tabular}{rr|rrr|rrr|rrr}
    \hline
 & & \multicolumn{3}{c|}{LR} & \multicolumn{3}{c|}{SLR} & \multicolumn{3}{c}{+\{All Cuts\}} \\
     \multicolumn{1}{c}{group}	& \multicolumn{1}{c|}{$n$} & \multicolumn{1}{c}{gap}	& \multicolumn{1}{c}{std}	& \multicolumn{1}{c|}{time}	&
     \multicolumn{1}{c}{gap} & \multicolumn{1}{c}{std}	& \multicolumn{1}{c|}{time}	&
     \multicolumn{1}{c}{gap} &	\multicolumn{1}{c}{std}	& \multicolumn{1}{c}{time} \\ 
     \hline
J50	&	540	&	0,750	&	0,309	&	0,6	&	0,749	&	0,309	&	6,4	&	\textbf{0,654}	&	0,367	&	9669,4	\\
J100	&	540	&	0,768	&	0,293	&	11,0	&	0,768	&	0,293	&	35,6	&	\textbf{0,688}	&	0,354	&	11536,6	\\
JAll50+	&	1616	&	0,586	&	0,250	&	6,6	&	0,584	&	0,277	&	30,1	&	\textbf{0,463}	&	0,227	&	13746,8	\\
Jall100+	&	1489	&	0,644	&	0,272	&	54,8	&	0,644	&	0,272	&	144,0	&	\textbf{0,530}	&	0,265	&	14221,9	\\ \hline
Total	&	4185	&	0,687	&	0,281	&	18,2	&	0,686	&	0,288	&	54,0	&	\textbf{0,584}	&	0,303	&	12293,7	\\
\hline
    \end{tabular}
    \label{tab:resultscpMMLIB} 
\end{table*}

The results show that the cutting plane was very effective for all benchmark datasets, improving the average values by approximately 15\%. As in earlier comparisons, the SLR slightly improves the integrality gaps for MMLIB instances too. However, 135 of the JA11 instances ran out of memory, which have been removed from the results presented in the table to provide a fair comparison.

\subsection{Branch \& Cut Experiments}

We now explore how the different cut types can be added dynamically in a Branch-and-Cut manner, to improve the solution process using the general purpose MILP solver Gurobi. Experiments were executed in order to compare the results achieved by solving the model with the inclusion of cuts into the root plus precedence cuts into a callback\footnote{a callback is a user function that is called periodically by the Gurobi optimizer in order to allow the user to query or modify the state of the optimization \cite{gurobi}} procedure when the lower bound is improved and the results achieved by solving the model without cuts just with the preprocessing input data and Gurobi cuts. 

Table \ref{tab:resultsbc} summarizes the integrality gaps\footnote{the integrality gaps is computed as in the previous experiments, since it was not possible to generate an incumbent solution for all instances, so it is not possible to obtain the optimality gap for some instances} (average and standard deviation), for open instances with $\alpha > 0$ from PSPLIB and MISTA. The experiments for our approach have been limited to 24 hours of computing time. For Gurobi cuts, the first experiments have been limited to 24 hours and the second have been limited to 48 hours, in order to ensure a fair comparison since we not consider the time spent to insert all cuts into the root node for our approach. Initial solutions available at MISTA website were inserted only for the A dataset.
\setlength\tabcolsep{1.5pt}
\begin{table*}[!ht]
    \footnotesize
    \centering
    \caption{Average and standard deviation for the integrality gaps obtained by the B\&C with all cuts at root plus LPR in the callback procedure}
    \begin{tabular}{rr|rr|rr|rr}
    \hline
     & & \multicolumn{4}{c}{Gurobi cuts} & \multicolumn{2}{|c}{ +\{Our approach\}}  \\ 
     & & \multicolumn{2}{c}{gap (24h)} & \multicolumn{2}{|c}{gap (48h)}  & \multicolumn{2}{|c}{gap (24h)} \\ 
     group	&	$n$	&	avg 	&	std	&	avg	&	std	&	avg 	&	std	\\  \hline 
A	&	10	&	0.210	&	0.165	&	 0.158 &	0.139 & \textbf{0.139}	& \textbf{0.124}	\\ 
J30	&	245	&	0.007	&	0.022	&	0.006	&	0.022 & \textbf{0.005}	& \textbf{0.020}	\\ 
J60	&	57	&	0.252	&	0.115	&	0.237	&	0.103 & \textbf{0.184}	& \textbf{0.084}  \\ 
J90	&	80	&	0.268	&	0.162	&	0.253	&	0.139 & \textbf{0.201}	& \textbf{0.084}	\\ 
J120	&	390	&	0.299	&	0.253	&	0.292	&	0.252 & \textbf{0.249}	& \textbf{0.224} \\	
\hline
   \end{tabular}
   \label{tab:resultsbc} 
\end{table*}

By analyzing Table \ref{tab:resultsbc} we can see that better results are obtained with the introduction of cuts into the root and with the LPR cut into the callback procedure. The average integrality gaps are reduced. Also, for some particular instances like $j12049\_8$, $j12021\_1$, $j12022\_8$, $j3037\_1$, $j3037\_7$, $j6046\_5$, $j6030\_2$, among others, the average optimality gap achieved within $24$ hours of computing time was quite low (0.09\%) when adding the cuts. Almost all open instances with $\alpha = 0$ from PSPLIB have been easily solved to optimality, except for some instances in set with $120$ jobs. 

We now explore how the two settings compare throughout the optimization process to find feasible solutions and to prove optimality. Table \ref{tab:resultsbceasy} summarizes the results for datasets from MISTA and PSPLIB for the three problem variants, also including those instances with $\alpha = 0$. Comparing to \citet{Toffolo2016}, we solve to optimality, for the first time, $1$ instance from \texttt{A} set.  Comparing to \citet{Schnell2017} we solve the \texttt{J30} set, for the first time, 13 instances to optimality, and prove infeasibility for 89 of the instances.  

\setlength\tabcolsep{1.5pt}
\begin{table*}[!ht]
    \footnotesize
    \centering
    \caption{Solutions obtained by the B\&C with all cuts at root plus LPR in the callback procedure}
    \begin{tabular}{rr|rrr|rrr}
    \hline
     & & \multicolumn{3}{c}{Gurobi cuts} & \multicolumn{3}{|c}{+\{Our approach\}} \\ 
     group	&	$n$	& opt &	fea & inf	&  opt & fea & inf	\\ \hline
     A	&  10	& \textbf{4} & 6 & 0 &  \textbf{4}  & 6 & 0 \\
     J30	&  641	& 530 &  6 & \textbf{89} &  \textbf{553}  & \textbf{15} & \textbf{89} \\
     J60	&	79	 & 24  &  0 &  0 &  \textbf{26}  &   \textbf{2} &  0\\ 
     J90	&  105	& \textbf{27}  &  0 &  0 &  \textbf{27}  &   0 &  0 \\
    J120	&  514	& 160 &  2 &  0 & \textbf{181}  &   \textbf{3} &  0 \\ \hline
   \end{tabular}
   \label{tab:resultsbceasy} 
\end{table*}

The results indicate that our approach has been able to prove optimality for more instances than using Gurobi without our cuts. For some specific instances the optimal value is found only when cuts are added to the model. In addition, Gurobi solver proves that some instances are infeasible. 

Table \ref{tab:resultsbcMMLIB} summarizes the results for our approach on the MMLIB datasets. The table shows the number of instances that have been solved to optimality (opt) and the number of instances for which a feasible solution has been found, but optimality has not been proven (fea). Further, the table shows the number of instances for which our solutions improve upon those reported by the website\footnote{http://mmlib.eu/solutions.php} (imp) and those reported in \cite{Schnell2017}, as well the number of instances for which optimality has been proven for the first time (new opt).

\setlength\tabcolsep{1.5pt}
\begin{table*}[!ht]
    \footnotesize
    \centering
    \caption{Solutions obtained by the B\&C with all cuts at root plus LPR in the callback procedure}
    \begin{tabular}{rr|rrrrr}
    \hline
     & & \multicolumn{5}{c}{all cuts at root + LPR callback} \\ 
     group	&	$n$	& opt &	new opt & fea & imp	& imp \cite{Schnell2017}  \\ \hline
J50 & 540 & 450 & 29 & 32 & 0 & 48  \\ 
J100 & 540 & 410 & 48 & 43 & 7 & 114  \\ 
JAll50+ & 1620  & 689 & 252 & 159 & 59 & 401  \\ 
JAll100+ & 1620  & 482 & 177 & 205 & 174  & 440  \\ \hline
   \end{tabular}
   \label{tab:resultsbcMMLIB} 
\end{table*}

\begin{figure*}[!ht]
	\centering
  \includegraphics[width=.8\linewidth]{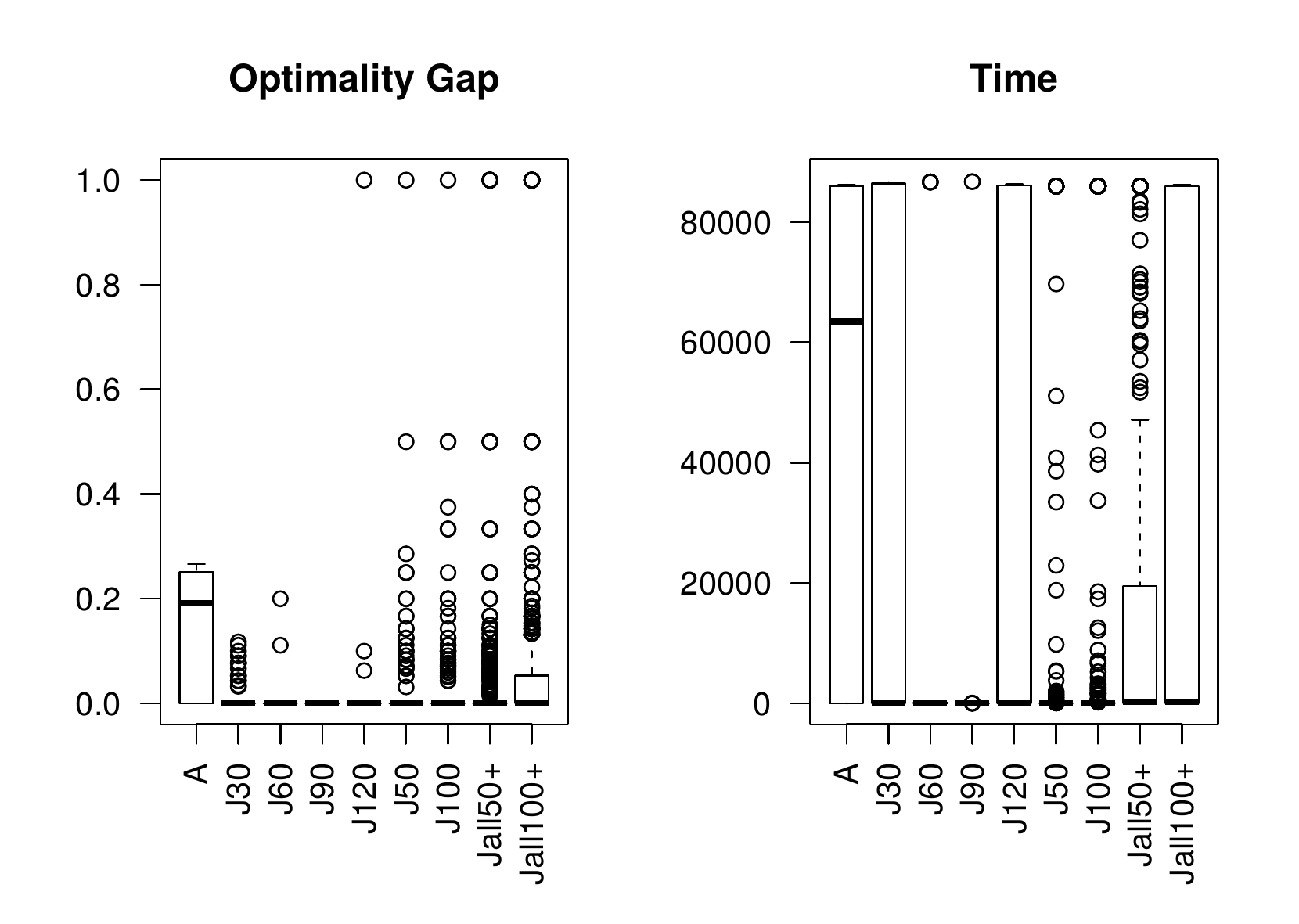}
   \caption{Boxplots with informations about optimality gap and computing time (sec.) of instances from PSPLIB, MISTA and MMLIB }
     \label{fig:BoxPlot01GapTime}  
\end{figure*}

Figure \ref{fig:BoxPlot01GapTime} presents boxplots of the optimality gaps and computing times for our B\&C approach for instances from PSPLIB, MISTA and MMLIB for which feasible or optimal solutions have been found.

 The results presented in the first figure suggest that the gap values equals to 0 are represented by the median in the box plot for almost all datasets, except for the A and Jall100+ datasets. The high outliers for the datasets J120, J50, J100, Jall50+ and Jall100+ indicate that the presented algorithms still require improvement in order to deliver robust results on all instances. Analyzing the computing times it can be observed that our approach quickly proves optimality for all but dataset A. Finally, we also note that some of the instances between the second and third quartile hit the time limit.

In summary, optimality was proven for the first time, for $247$ instances from PSPLIB, for $1$ instance from MISTA and for $506$ instances from MMLIB, totalizing $754$ instances. The LP and solution files for all benchmark datasets are available for download at \texttt{http://professor.ufop.br/janniele/downloads}.

\section{Conclusion and Future Works}
\label{sec:conclusion}
In this paper, new mixed-integer linear programming based methods were proposed to improve the linear programming relaxation of compact formulations for the Resource Constrained Project Scheduling Problem. All methods were extensively evaluated in three RCPSP variants.

An effective preprocessing procedure to strengthen renewable resources constraints was devised. This procedure was capable of improving the lower bounds produced at the root node without any increase in the size of the linear programs.
	
A parallel cutting plane algorithm was developed including five families of cuts: lifted precedence and cover cuts, cliques, odd-holes and strengthened Chv\'atal-Gomory cuts. A dense conflict graph, considering feasibility and optimality conditions, was created at each iteration and used by these cut generators in strengthening procedures. All cuts contributed to improve the lower bounds, specially when they are together in the cutting plane. However, the lifted precedence cuts were the most effective for all variants. The strengthened Chv\'atal-Gomory cuts were specially effective in a group of multi-project instances. These results indicate that an instance feature based tuning of the cut generators may be beneficial.

With the improved linear programming formulations produced with our methods, $754$ open instances from literature were solved for the first time: $247$ instances from PSPLIB, $1$ instance from the MISTA Challenge and for $506$ instances from MMLIB.

The method still has room for improvement. These techniques could be hybridized with constraint propagation, as proposed in previous studies. The separation of some inequalities, such as the strengthened Chv\'atal-Gomory cuts and the re-optimization of the large linear programs are still quite expensive. The continuous improvements in future MILP solvers will likely speed up these two steps. Finally, the development of a complete B\&C with improved branch and node selection strategies for the RCPSP, including our preprocessing and cutting planes routines, is also a promising future path.
	
\bigskip
\noindent \textbf{Acknowledgements.} The authors thank UFOP, CNPq (student with scholarship - Brazil), FAPEMIG, and Compute Canada for supporting this research.





\bigskip
\noindent \textbf{References} 
    \bibliographystyle{elsarticle-harv} 
    \bibliography{rcpsp-preproc-and-cutting-planes.bib}

\begin{thebibliography}{49}
\expandafter\ifx\csname natexlab\endcsname\relax\def\natexlab#1{#1}\fi
\providecommand{\url}[1]{\texttt{#1}}
\providecommand{\href}[2]{#2}
\providecommand{\path}[1]{#1}
\providecommand{\DOIprefix}{doi:}
\providecommand{\ArXivprefix}{arXiv:}
\providecommand{\URLprefix}{URL: }
\providecommand{\Pubmedprefix}{pmid:}
\providecommand{\doi}[1]{\href{http://dx.doi.org/#1}{\path{#1}}}
\providecommand{\Pubmed}[1]{\href{pmid:#1}{\path{#1}}}
\providecommand{\bibinfo}[2]{#2}
\ifx\xfnm\relax \def\xfnm[#1]{\unskip,\space#1}\fi
\bibitem[{Applegate and Cook(1991)}]{Applegate1991}
\bibinfo{author}{Applegate, D.}, \bibinfo{author}{Cook, W.},
  \bibinfo{year}{1991}.
\newblock \bibinfo{title}{A computational study of the job-shop scheduling
  problem}.
\newblock \bibinfo{journal}{ORSA Journal on Computing} \bibinfo{volume}{3},
  \bibinfo{pages}{149--1556}.
\bibitem[{Artigues(2017)}]{Artigues2017}
\bibinfo{author}{Artigues, C.}, \bibinfo{year}{2017}.
\newblock \bibinfo{title}{On the strength of time-indexed formulations for the
  resource-constrained project scheduling problem}.
\newblock \bibinfo{journal}{Operations Research Letters} \bibinfo{volume}{45},
  \bibinfo{pages}{154--–159}.
\bibitem[{Artigues et~al.(2008)Artigues, Demassey and
  N{\'{e}}on}]{Artigues2008}
\bibinfo{author}{Artigues, C.}, \bibinfo{author}{Demassey, S.},
  \bibinfo{author}{N{\'{e}}on, E.}, \bibinfo{year}{2008}.
\newblock \bibinfo{title}{{Resource-Constrained Project Scheduling: Models,
  Algorithms, Extensions and Applications}}.
\newblock \bibinfo{publisher}{{I}{S}{T}{E} Ltd and John Wiley \& Sons, Inc}.
\bibitem[{Atamt{\"{u}}rk et~al.(2000)Atamt{\"{u}}rk, Nemhauser and
  Savelsbergh}]{atamturk2000}
\bibinfo{author}{Atamt{\"{u}}rk, A.}, \bibinfo{author}{Nemhauser, G.L.},
  \bibinfo{author}{Savelsbergh, M.W.P.}, \bibinfo{year}{2000}.
\newblock \bibinfo{title}{{Conflict graphs in solving integer programming
  problems}}.
\newblock \bibinfo{journal}{European Journal of Operational Research}
  \bibinfo{volume}{121}, \bibinfo{pages}{40--55}.
\bibitem[{Balas(1975)}]{Balas1975}
\bibinfo{author}{Balas, E.}, \bibinfo{year}{1975}.
\newblock \bibinfo{title}{{Facets of the knapsack polytope}}.
\newblock \bibinfo{journal}{Mathematical Programming} \bibinfo{volume}{8},
  \bibinfo{pages}{146--164}.
\bibitem[{Balas and Zemel(1978)}]{Balas1978}
\bibinfo{author}{Balas, E.}, \bibinfo{author}{Zemel, E.}, \bibinfo{year}{1978}.
\newblock \bibinfo{title}{Facets of the knapsack polytope from minimal covers}.
\newblock \bibinfo{journal}{SIAM Journal on Applied Mathematics}
  \bibinfo{volume}{34}, \bibinfo{pages}{119--148}.
\bibitem[{Baptiste and Demassey(2004)}]{Baptiste2004}
\bibinfo{author}{Baptiste, P.}, \bibinfo{author}{Demassey, S.},
  \bibinfo{year}{2004}.
\newblock \bibinfo{title}{Tight lp bounds for resource constrained project
  scheduling}.
\newblock \bibinfo{journal}{OR Spectrum} \bibinfo{volume}{26},
  \bibinfo{pages}{251--262}.
\bibitem[{Blazewicz et~al.(1983)Blazewicz, Lenstra and {Rinnooy
  Kan}}]{Blazewicz1983}
\bibinfo{author}{Blazewicz, J.}, \bibinfo{author}{Lenstra, J.},
  \bibinfo{author}{{Rinnooy Kan}, A.}, \bibinfo{year}{1983}.
\newblock \bibinfo{title}{Scheduling subject to resource constraints:
  classification and complexity}.
\newblock \bibinfo{journal}{Discrete Appl. Math} \bibinfo{volume}{5},
  \bibinfo{pages}{11--24}.
\bibitem[{Boyd(1992)}]{Boyd1992}
\bibinfo{author}{Boyd, E.}, \bibinfo{year}{1992}.
\newblock \bibinfo{title}{Fenchel cutting planes for integer programming}.
\newblock \bibinfo{journal}{Operations Research} \bibinfo{volume}{42},
  \bibinfo{pages}{53--64}.
\bibitem[{Boyd(1994)}]{Boyd1994}
\bibinfo{author}{Boyd, E.}, \bibinfo{year}{1994}.
\newblock \bibinfo{title}{Solving 0/1 integer programs with enumeration cutting
  planes}.
\newblock \bibinfo{journal}{Annals of Operations Research}
  \bibinfo{volume}{50}, \bibinfo{pages}{61--72}.
\bibitem[{Brito et~al.(2015)Brito, Santos and Poggi}]{Brito2015}
\bibinfo{author}{Brito, S.S.}, \bibinfo{author}{Santos, H.G.},
  \bibinfo{author}{Poggi, M.}, \bibinfo{year}{2015}.
\newblock \bibinfo{title}{A computational study of conflict graphs and
  aggressive cut separation in integer programming}.
\newblock \bibinfo{journal}{Electronic Notes in Discrete Mathematics}
  \bibinfo{volume}{50}, \bibinfo{pages}{355--360}.
\bibitem[{Brucker et~al.(1998)Brucker, Knust, Schoo and
  Thiele}]{Brucker1998272}
\bibinfo{author}{Brucker, P.}, \bibinfo{author}{Knust, S.},
  \bibinfo{author}{Schoo, A.}, \bibinfo{author}{Thiele, O.},
  \bibinfo{year}{1998}.
\newblock \bibinfo{title}{A branch and bound algorithm for the
  resource-constrained project scheduling problem}.
\newblock \bibinfo{journal}{European Journal of Operational Research}
  \bibinfo{volume}{107}, \bibinfo{pages}{272--288}.
\bibitem[{Cavalcante et~al.(2001)Cavalcante, de~Souza, Savelsbergh, Y and
  Wolsey}]{Cavalcante2001}
\bibinfo{author}{Cavalcante, C.}, \bibinfo{author}{de~Souza, C.},
  \bibinfo{author}{Savelsbergh, M.}, \bibinfo{author}{Y, W.},
  \bibinfo{author}{Wolsey, L.A.}, \bibinfo{year}{2001}.
\newblock \bibinfo{title}{cheduling projects with labor constraints}.
\newblock \bibinfo{journal}{Discrete Applied Mathematics}
  \bibinfo{volume}{112}, \bibinfo{pages}{27--52}.
\bibitem[{Chakrabortty et~al.(2015)Chakrabortty, Sarker and Essam}]{Ripon2015}
\bibinfo{author}{Chakrabortty, R.K.}, \bibinfo{author}{Sarker, R.A.},
  \bibinfo{author}{Essam, D.L.}, \bibinfo{year}{2015}.
\newblock \bibinfo{title}{Resource constrained project scheduling: A branch and
  cut approach}, in: \bibinfo{booktitle}{International Conference on Computers
  and Industrial Engineering}, pp. \bibinfo{pages}{552--559}.
\bibitem[{Christofides et~al.(1987)Christofides, Alvarez-Valdes and
  Tamarit}]{CHRISTOFIDES1987262}
\bibinfo{author}{Christofides, N.}, \bibinfo{author}{Alvarez-Valdes, R.},
  \bibinfo{author}{Tamarit, J.}, \bibinfo{year}{1987}.
\newblock \bibinfo{title}{Project scheduling with resource constraints: A
  branch and bound approach}.
\newblock \bibinfo{journal}{European Journal of Operational Research}
  \bibinfo{volume}{29}, \bibinfo{pages}{262--273}.
\bibitem[{Chv\'atal(1973)}]{Chvatal1973}
\bibinfo{author}{Chv\'atal, V.}, \bibinfo{year}{1973}.
\newblock \bibinfo{title}{Edmonds polytopes and a hierarchy of combinatorial
  problems}.
\newblock \bibinfo{journal}{Discrete Mathematics} \bibinfo{volume}{4},
  \bibinfo{pages}{305--–337}.
\bibitem[{Demassey et~al.(2005)Demassey, Artigues and Michelon}]{Demassey2005}
\bibinfo{author}{Demassey, S.}, \bibinfo{author}{Artigues, C.},
  \bibinfo{author}{Michelon, P.}, \bibinfo{year}{2005}.
\newblock \bibinfo{title}{Constraint propagation based cutting planes : an
  application to the resource-constrained project scheduling problem}.
\newblock \bibinfo{journal}{INFORMS Journal on Computing} \bibinfo{volume}{17},
  \bibinfo{pages}{52--65}.
\bibitem[{Demeulemeester and Herroelen(1992)}]{Demeulemeester1992}
\bibinfo{author}{Demeulemeester, E.}, \bibinfo{author}{Herroelen, W.},
  \bibinfo{year}{1992}.
\newblock \bibinfo{title}{Recent advances in branch-and-bound procedures for
  resource-constrained project scheduling problems}, in:
  \bibinfo{booktitle}{Paper presented at the Summer School on Scheduling Theory
  and Its Applications}, pp. \bibinfo{pages}{1--32}.
\bibitem[{Demeulemeester and Herroelen(2002)}]{Deumeulemeester2002}
\bibinfo{author}{Demeulemeester, E.L.}, \bibinfo{author}{Herroelen, W.S.},
  \bibinfo{year}{2002}.
\newblock \bibinfo{title}{Project Scheduling: A Research Handbook}.
\newblock \bibinfo{publisher}{Kluwer Academic Publishers}.
\bibitem[{Fischetti and Lodi(2007)}]{Fischetti2007}
\bibinfo{author}{Fischetti, M.}, \bibinfo{author}{Lodi, A.},
  \bibinfo{year}{2007}.
\newblock \bibinfo{title}{Optimizing over the first {C}hv{\'a}tal closure}.
\newblock \bibinfo{journal}{Mathematical Programming} \bibinfo{volume}{110},
  \bibinfo{pages}{3--20}.
\bibitem[{Garey and Johnson(1979)}]{Garey1979}
\bibinfo{author}{Garey, M.R.}, \bibinfo{author}{Johnson, D.S.},
  \bibinfo{year}{1979}.
\newblock \bibinfo{title}{Computers and Intractability: A Guide to the Theory
  of NP-Completeness}.
\newblock \bibinfo{publisher}{W. H. Freeman \& Co.}, \bibinfo{address}{New
  York, NY, USA}.
\bibitem[{Gu et~al.(2000)Gu, Nemhauser and Savelsbergh}]{Gu2000}
\bibinfo{author}{Gu, Z.}, \bibinfo{author}{Nemhauser, G.},
  \bibinfo{author}{Savelsbergh, M.}, \bibinfo{year}{2000}.
\newblock \bibinfo{title}{Sequence independent lifting in mixed integer
  programming}.
\newblock \bibinfo{journal}{Journal of Combinatorial Optimization}
  \bibinfo{volume}{4}, \bibinfo{pages}{109--129}.
\bibitem[{Gurobi~Optimization(2016)}]{gurobi}
\bibinfo{author}{Gurobi~Optimization, I.}, \bibinfo{year}{2016}.
\newblock \bibinfo{title}{Gurobi optimizer: Reference manual}.
\newblock \URLprefix \url{http://www.gurobi.com/documentation/7.0/refman.pdf}.
\bibitem[{Hardin et~al.(2008)Hardin, Nemhauser and Savelsbergh}]{Hardin2008}
\bibinfo{author}{Hardin, J.R.}, \bibinfo{author}{Nemhauser, G.L.},
  \bibinfo{author}{Savelsbergh, M.W.}, \bibinfo{year}{2008}.
\newblock \bibinfo{title}{Strong valid inequalities for the
  resource-constrained scheduling problem with uniform resource requirements}.
\newblock \bibinfo{journal}{Discrete Optimization} \bibinfo{volume}{5},
  \bibinfo{pages}{19--35}.
\bibitem[{Johnson et~al.(1985)Johnson, Kostreva and Suhl}]{Johnson1985}
\bibinfo{author}{Johnson, E.}, \bibinfo{author}{Kostreva, M.},
  \bibinfo{author}{Suhl, U.}, \bibinfo{year}{1985}.
\newblock \bibinfo{title}{Solving 0-1 integer programming problems arising from
  large scale planning models}.
\newblock \bibinfo{journal}{Operations Research} \bibinfo{volume}{33},
  \bibinfo{pages}{803--819}.
\bibitem[{{Kelley Jr} and Walker(1959)}]{Kelley1959}
\bibinfo{author}{{Kelley Jr}, J.}, \bibinfo{author}{Walker, M.R.},
  \bibinfo{year}{1959}.
\newblock \bibinfo{title}{Critical-path planning and scheduling}, in:
  \bibinfo{booktitle}{Eastern Joint IRE-AIEE-ACM Computer Conference},
  \bibinfo{publisher}{ACM}. pp. \bibinfo{pages}{160--173}.
\bibitem[{Kolisch(1995)}]{Kolisch1995}
\bibinfo{author}{Kolisch, R.}, \bibinfo{year}{1995}.
\newblock \bibinfo{title}{Project Scheduling under Resource Constraints:
  Efficient Heuristics for Several Problem Classes}.
\newblock Number \bibinfo{number}{IX, 212} in \bibinfo{series}{Production and
  Logistics}. \bibinfo{edition}{1} ed., \bibinfo{publisher}{Physica-Verlag
  Heidelberg}.
\bibitem[{Kolisch and Sprecher(1996)}]{Kolisch1996}
\bibinfo{author}{Kolisch, R.}, \bibinfo{author}{Sprecher, A.},
  \bibinfo{year}{1996}.
\newblock \bibinfo{title}{{P}{S}{P}{L}{I}{B} - a project scheduling problem
  library}.
\newblock \bibinfo{journal}{European Journal of Operational Research}
  \bibinfo{volume}{96}, \bibinfo{pages}{205--216}.
\bibitem[{Kone et~al.(2011)Kone, Artigues, Lopez and Mongeau}]{Kone2011}
\bibinfo{author}{Kone, O.}, \bibinfo{author}{Artigues, C.},
  \bibinfo{author}{Lopez, P.}, \bibinfo{author}{Mongeau, M.},
  \bibinfo{year}{2011}.
\newblock \bibinfo{title}{Event-based {M}{I}{L}{P} models for
  resource-constrained project scheduling problems}.
\newblock \bibinfo{journal}{Computers and Operations Research}
  \bibinfo{volume}{38}, \bibinfo{pages}{3--13}.
\bibitem[{Land and Doig(2010)}]{Land2010}
\bibinfo{author}{Land, A.H.}, \bibinfo{author}{Doig, A.G.},
  \bibinfo{year}{2010}.
\newblock \bibinfo{title}{An Automatic Method for Solving Discrete Programming
  Problems}. \bibinfo{publisher}{Springer Berlin Heidelberg},
  \bibinfo{address}{Berlin, Heidelberg}.
\newblock pp. \bibinfo{pages}{105--132}.
\bibitem[{Letchford et~al.(2016)Letchford, Marzi, Rossi and
  Smriglio}]{Letchford2016}
\bibinfo{author}{Letchford, A.N.}, \bibinfo{author}{Marzi, F.},
  \bibinfo{author}{Rossi, F.}, \bibinfo{author}{Smriglio, S.},
  \bibinfo{year}{2016}.
\newblock \bibinfo{title}{{Strengthening Chv{\'{a}}tal-Gomory Cuts for the
  Stable Set Problem}}, in: \bibinfo{booktitle}{Combinatorial Optimization},
  \bibinfo{publisher}{Springer International Publishing}. pp.
  \bibinfo{pages}{201--212}.
\bibitem[{Liu et~al.(2014)Liu, Shan and Wu}]{Liu2014}
\bibinfo{author}{Liu, M.}, \bibinfo{author}{Shan, M.}, \bibinfo{author}{Wu,
  J.}, \bibinfo{year}{2014}.
\newblock \bibinfo{title}{Multiple {R}\&{D} projects scheduling optimization
  with improved particle swarm algorithm}.
\newblock \bibinfo{journal}{The Scientific World Journal}
  \bibinfo{volume}{2014}, \bibinfo{pages}{652135}.
\bibitem[{Liu and Wang(2008)}]{Liu2008}
\bibinfo{author}{Liu, S.S.}, \bibinfo{author}{Wang, C.J.},
  \bibinfo{year}{2008}.
\newblock \bibinfo{title}{Resource-constrained construction project scheduling
  model for profit maximization considering cash flow}.
\newblock \bibinfo{journal}{Automation in Construction} \bibinfo{volume}{17},
  \bibinfo{pages}{966--974}.
\bibitem[{Mingozzi et~al.(1998)Mingozzi, Maniezzo, Ricciardelli and
  Bianco}]{Mingozzi1998}
\bibinfo{author}{Mingozzi, A.}, \bibinfo{author}{Maniezzo, V.},
  \bibinfo{author}{Ricciardelli, S.}, \bibinfo{author}{Bianco, L.},
  \bibinfo{year}{1998}.
\newblock \bibinfo{title}{{An Exact Algorithm for the Resource-Constrained
  Project Scheduling Problem Based on a New Mathematical Formulation}}.
\newblock \bibinfo{journal}{Management Science} \bibinfo{volume}{44},
  \bibinfo{pages}{714--729}.
\bibitem[{Nemhauser and Vance(1994)}]{Nemhauser1994}
\bibinfo{author}{Nemhauser, G.L.}, \bibinfo{author}{Vance, P.H.},
  \bibinfo{year}{1994}.
\newblock \bibinfo{title}{{Lifted cover facets of the 0–1 knapsack polytope
  with GUB constraints}}.
\newblock \bibinfo{journal}{Operations Research Letters} \bibinfo{volume}{16},
  \bibinfo{pages}{255--263}.
\bibitem[{Padberg(1973)}]{Padberg1973}
\bibinfo{author}{Padberg, M.W.}, \bibinfo{year}{1973}.
\newblock \bibinfo{title}{On the facial structure of set packing polyhedra}.
\newblock \bibinfo{journal}{Mathematical Programming} \bibinfo{volume}{5},
  \bibinfo{pages}{199--215}.
\bibitem[{Patterson and Huber(1974)}]{Patterson1974}
\bibinfo{author}{Patterson, J.}, \bibinfo{author}{Huber, W.},
  \bibinfo{year}{1974}.
\newblock \bibinfo{title}{A horizon-varying zero-one approach to project
  scheduling}.
\newblock \bibinfo{journal}{Management Science} \bibinfo{volume}{20},
  \bibinfo{pages}{990--998}.
\bibitem[{Pritsker et~al.(1969)Pritsker, Watters and Wolfe}]{Pritsker1969}
\bibinfo{author}{Pritsker, A.}, \bibinfo{author}{Watters, L.},
  \bibinfo{author}{Wolfe, P.}, \bibinfo{year}{1969}.
\newblock \bibinfo{title}{Multi project scheduling with limited resources: A
  zero-one programming approach}.
\newblock \bibinfo{journal}{Management Science} \bibinfo{volume}{3416},
  \bibinfo{pages}{93--108}.
\bibitem[{Riedler et~al.(2020)Riedler, Jatschka, Maschler and
  Raidl}]{Riedler2017}
\bibinfo{author}{Riedler, M.}, \bibinfo{author}{Jatschka, T.},
  \bibinfo{author}{Maschler, J.}, \bibinfo{author}{Raidl, G.R.},
  \bibinfo{year}{2020}.
\newblock \bibinfo{title}{An iterative time-bucket refinement algorithm for a
  high resolution resource-constrained project scheduling problem}.
\newblock \bibinfo{journal}{International Transactions in Operational Research}
  \bibinfo{volume}{27}, \bibinfo{pages}{573--613}.
\bibitem[{Sankaran et~al.(1999)Sankaran, Bricker,  and Juang}]{Sankaran1999}
\bibinfo{author}{Sankaran, J.K.}, \bibinfo{author}{Bricker, D.L.}, ,
  \bibinfo{author}{Juang, S.}, \bibinfo{year}{1999}.
\newblock \bibinfo{title}{A strong fractional cutting plane algorithm for
  resource-constrained project scheduling}.
\newblock \bibinfo{journal}{International Journal of Industrial Engineering -
  Applications and Practice} \bibinfo{volume}{6}, \bibinfo{pages}{99--111}.
\bibitem[{Santos et~al.(2016)Santos, Toffolo, Gomes and Ribas}]{Santos2016}
\bibinfo{author}{Santos, H.G.}, \bibinfo{author}{Toffolo, T.A.M.},
  \bibinfo{author}{Gomes, R.A.M.}, \bibinfo{author}{Ribas, S.},
  \bibinfo{year}{2016}.
\newblock \bibinfo{title}{Integer programming techniques for the nurse
  rostering problem}.
\newblock \bibinfo{journal}{Annals of Operations Research}
  \bibinfo{volume}{239}, \bibinfo{pages}{225--251}.
\bibitem[{Schnell and Hartl(2017)}]{Schnell2017}
\bibinfo{author}{Schnell, A.}, \bibinfo{author}{Hartl, R.F.},
  \bibinfo{year}{2017}.
\newblock \bibinfo{title}{{On the generalization of constraint programming and
  boolean satisfiability solving techniques to schedule a resource-constrained
  project consisting of multi-mode jobs}}.
\newblock \bibinfo{journal}{Operations Research Perspectives}
  \bibinfo{volume}{4}, \bibinfo{pages}{1--11}.
\bibitem[{Schwindt and Zimmermann(2015)}]{Schwindt2015}
\bibinfo{author}{Schwindt, C.}, \bibinfo{author}{Zimmermann, J.},
  \bibinfo{year}{2015}.
\newblock \bibinfo{title}{Handbook on Project Management and Scheduling}.
  volume~\bibinfo{volume}{1} of \textit{\bibinfo{series}{International
  Handbooks on Information Systems}}.
\newblock \bibinfo{publisher}{Springer International Publishing}.
\bibitem[{de~Souza and Wolsey(1997)}]{deSouza1997}
\bibinfo{author}{de~Souza, C.C.}, \bibinfo{author}{Wolsey, L.A.},
  \bibinfo{year}{1997}.
\newblock \bibinfo{title}{Scheduling projects with labour constraints}.
\newblock \bibinfo{journal}{Technical report, Instituto de Computação,
  Universidade Estadual de Campinas} .
\bibitem[{Toffolo et~al.(2016)Toffolo, Santos, Carvalho and
  Soares}]{Toffolo2016}
\bibinfo{author}{Toffolo, T.}, \bibinfo{author}{Santos, H.},
  \bibinfo{author}{Carvalho, M.}, \bibinfo{author}{Soares, J.},
  \bibinfo{year}{2016}.
\newblock \bibinfo{title}{An integer programming approach to the multimode
  resource-constrained multiproject scheduling problem}.
\newblock \bibinfo{journal}{Journal of Scheduling} \bibinfo{volume}{19},
  \bibinfo{pages}{295--307}.
\bibitem[{Van~Peteghem and Vanhoucke(2014)}]{VanPeteghem2014}
\bibinfo{author}{Van~Peteghem, V.}, \bibinfo{author}{Vanhoucke, M.},
  \bibinfo{year}{2014}.
\newblock \bibinfo{title}{An experimental investigation of metaheuristics for
  the multi-mode resource-constrained project scheduling problem on new dataset
  instances}.
\newblock \bibinfo{journal}{European Journal of Operational Research}
  \bibinfo{volume}{235}, \bibinfo{pages}{62--72}.
\bibitem[{Wauters et~al.(2016)Wauters, Kinable, Smet, Vancroonenburg, {Vanden
  Berghe} and Verstichel}]{Wauters2016}
\bibinfo{author}{Wauters, T.}, \bibinfo{author}{Kinable, J.},
  \bibinfo{author}{Smet, P.}, \bibinfo{author}{Vancroonenburg, W.},
  \bibinfo{author}{{Vanden Berghe}, G.}, \bibinfo{author}{Verstichel, J.},
  \bibinfo{year}{2016}.
\newblock \bibinfo{title}{The multi-mode resource-constrained multi-project
  scheduling problem}.
\newblock \bibinfo{journal}{Journal of Scheduling} \bibinfo{volume}{19},
  \bibinfo{pages}{271--283}.
\bibitem[{Wolsey(1998)}]{Wolsey1998}
\bibinfo{author}{Wolsey, L.}, \bibinfo{year}{1998}.
\newblock \bibinfo{title}{Integer Programming}.
\newblock \bibinfo{publisher}{Wiley - Interscience series in discrete
  mathematics and optimization}.
\bibitem[{Zhu et~al.(2006)Zhu, Bard and Yu}]{Zhu2006}
\bibinfo{author}{Zhu, G.}, \bibinfo{author}{Bard, J.}, \bibinfo{author}{Yu,
  G.}, \bibinfo{year}{2006}.
\newblock \bibinfo{title}{A branch-and-cut procedure for the multimode
  resource-constrained project-scheduling problem}.
\newblock \bibinfo{journal}{INFORMS Journal on Computing} \bibinfo{volume}{18},
  \bibinfo{pages}{377--390}.

\end{thebibliography}





\end{document}